\documentclass[12pt]{article}
\usepackage{upgreek}
\usepackage[mathscr]{euscript}
\usepackage{amsmath,environ}
\usepackage{amsfonts} 
\usepackage{mathrsfs}
\usepackage{calrsfs}
\usepackage{graphicx,psfrag,color}
\usepackage{enumerate}

\newtheorem{theorem}{\bf Theorem}
\newtheorem{remark}[theorem]{\bf Remark}

\usepackage{xcolor}
\usepackage[giveninits=true,sorting=none]{biblatex}
\usepackage[colorlinks=true,linkcolor=brown, citecolor=blue, urlcolor=purple]{hyperref}
\usepackage[left]{lineno}
\usepackage{lipsum} 
\setpagewiselinenumbers
\usepackage{enumerate}
\newcommand{\la}{\mathrm{b}}
\NewEnviron{eqn}{\begin{equation}\begin{split} \BODY\end{split}\end{equation}}
\usepackage{MnSymbol}
\DeclareMathSymbol{\gimel}{\mathord}{MnSyC}{"B1}
\newcommand{\od}[2]{\frac{d#1}{d#2}}
\newcommand{\su}{\mathtt{u}}
\newcommand{\beqans}{\begin{subequations}\begin{eqnarray}}
\newcommand{\eeqans}[1]{\end{eqnarray}\label{#1}\end{subequations}}
\newcommand{\beqan}{\begin{eqnarray}}
\newcommand{\eeqan}{\end{eqnarray}}
\DeclareMathOperator{\totwave}{t}
\DeclareMathOperator{\inc}{i}
\DeclareMathOperator{\refwave}{r}
\DeclareMathOperator{\scawave}{s}
\DeclareMathOperator{\icf}{\omega}
\newcommand{\sv}{\mathtt{v}}
\DeclareMathOperator{\sQ}{\mathtt{Q}}
\DeclareMathOperator{\sH}{\mathtt{H}}
\DeclareMathOperator{\sR}{\mathtt{R}}
\DeclareMathOperator{\sh}{\mathtt{h}}
\DeclareMathOperator{\sq}{\mathtt{q}}
\DeclareMathOperator{\sr}{\mathtt{r}}
\newcommand*{\bfrac}[2]{\genfrac{}{}{0pt}{}{\raisebox{-.3em}{\scriptsize$#1$}}{\raisebox{.4em}{\scriptsize$#2$}}}
\DeclareMathOperator{\mathscrpring}{\mathrm{w}}
\numberwithin{equation}{section}

\hoffset=0pt\voffset=10pt\textwidth=500pt\textheight=700pt\topmargin=-80pt\oddsidemargin=-20pt\evensidemargin=-20pt\let\oldmarginpar\marginpar\renewcommand\marginpar[1]{\-\oldmarginpar[\raggedleft\footnotesize #1]{\raggedright\footnotesize #1}}

\begin{document}

\title{Discrete scattering by a pair of parallel defects\thanks{``Discrete scattering by a pair of parallel defects'', Philosophical Transactions of the Royal Society A: Mathematical, Physical and Engineering Sciences, 2019, Vol 378, 1--20 DOI 10.1098/rsta.2019.0102}}
\author{Basant Lal Sharma\thanks{Department of Mechanical Engineering, Indian Institute of Technology Kanpur, Kanpur, U. P. 208016, India ({bls@iitk.ac.in})}\\
\and
Gaurav Maurya\thanks{Department of Mechanical Engineering, Indian Institute of Technology Kanpur, Kanpur, U. P. 208016, India ({gmaurya@iitk.ac.in}).}
}
\date{}
\maketitle

\begin{abstract}
Scattering of a time harmonic anti-plane shear wave due to either a pair of crack tips or a pair of rigid constraint tips on square lattice is considered. 
The two problems correspond to the so called zero-offset case of 
scattering due to a pair of identical Sommerfeld screens. 
The peculiar structural symmetry allows the reduction of coupled equations to two scalar Wiener--Hopf equations and a total of four geometrically reduced problems on {lattice} half-plane.
Exact solution of each 
problem for incidence from the bulk lattice, as well as from an associated lattice waveguide, is constructed. 
A suitable superposition of the four expressions is used to construct the solution of the main 
problem.
The discrete paradigm involving the wave mode incident from the waveguide is relevant for modern applications where an investigation of mechanisms of electronic and thermal 
transport at nanoscale remains an interesting problem. 
\end{abstract}

\section{Introduction}
\label{intro}
A 
square lattice based analogue of 
a canonical problem in 
scattering theory \cite{cheney1951diffraction,Jones1,meister1996factorization,thompson2005mode} is discussed: a time harmonic lattice wave is incident upon a pair of semi-infinite parallel rows with either Neumann or Dirichlet condition.
It is instructive to recall that,
within the 
well established continuum framework,
the scattering problem 
finds relevance in 
electro-magnetism, acoustics, and allied subjects \cite{williams_1954,jones_1952,jull1973aperture,johansen1965radiation,crease1958propagation,Abrahams1,james1979double,kapoulitsas1984propagation,michaeli1985new,michaeli1996asymptotic}, as well as from the viewpoint of geometric and asymptotic approximations \cite{bowman1970comparison,BoersmaLee,MenendezLee}. 
Strikingly, in the presence of an offset between the 
edges, so called staggered case, the scattering problem is difficult to solve 
\cite{jones1973double,Jones3planes,Abrahams0} owing to
the complexity of matrix Wiener--Hopf ({{WH}}) factorization \cite{Heinslim,MeisterRottbrand,Meistersys1,Meistersys2,daniele1984solution,AbrahamsExpo}. 
On the other hand,
when the edges are not staggered an exact solution is well known \cite{Heins1,Heins2}; 
this 
also plays a crucial role for solving the problem with small stagger in light of an asymptotic technique \cite{Mishuris2014}. 

Within the discrete framework, the two structures, that is, a pair of 
parallel cracks (Neumann condition) or a pair of rigid constraints (Dirichlet condition), can be construed as the {two dimensional formulation} of a three dimensional structure with a pair of parallel atomically thin cracks or rigid inclusions. 
The latter can be envisaged for a crystal lattice having a symmetry that allows square sub-lattice planes and at the same time admits an out-of-plane displacement relative to such sub-lattices. 
Both cracks or rigid inclusions can extend indefinitely in one direction and are spaced apart by certain multiples of the lattice parameter. 
{In this paper, the incident lattice wave field as well as the scattered wave field is time harmonic with the the same frequency. Moreover, it is assumed that there is a very small amount of damping present in the medium which results into a complex valued frequency with vanishingly small but positive imaginary part.}
The angle of incidence of the incident wave
and the (real part of) incident wave frequency 
can be arbitrary chosen {according to} the passband of the {considered square lattice} structure \cite{Brillouin}.

\begin{figure}[htb!]
\centering
{\includegraphics[width=\textwidth]{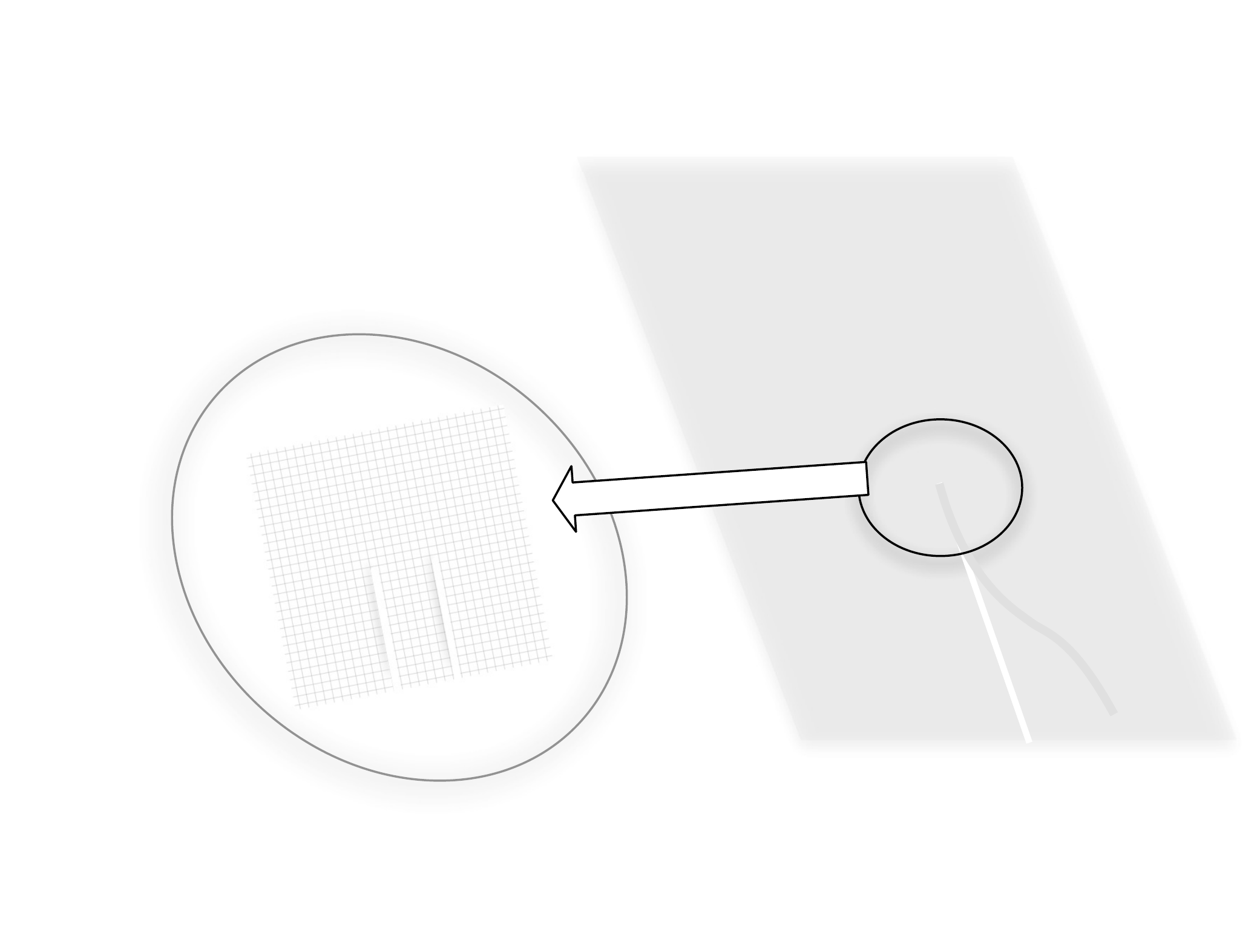}}
\caption{Square lattice attached to a single lead 
created by breaking bonds in a pair of semi-infinite rows.}
\label{sq_strip_heatbath_phonons}
\end{figure}

The present paper provides an exact solution of the stated discrete scattering problem and develops a far-field approximation for the incidence from the bulk lattice as well as for the incidence from the lattice waveguide formed between 
defects. Analytical expressions are also provided for certain physically relevant quantities, such the crack opening displacement, namely, the foremost (broken) bond length in any of the two cracks, and the displacement of a site adjacent to the rigid constraint tip.
In the scenario presented so far, the scattering problem attended in the paper 
involves a purely mechanical framework, however, there is a quantum-mechanical analogue as well within the tight binding approximation for the electronic wave function (see \S7.3 of \cite{Bls9hx} for honeycomb lattice). 
The mathematical connection between a specific lattice wave (phonon) based expression \cite{Bls9s} and that for the electronic wave has been spotted in this context \cite{Bls5c_tube, Bls5c_tube_media}.
In the mechanical framework, a significant scientific problem of current interest concerns the nature of energy transport in structures at small scales \cite{Cahill1}. In this regime, the transport is typically defined in terms of reflection and transmission, i.e., by so called the {\em Landauer viewpoint} \cite{Landauer1957,Landauer1992,Imry1999}. 
The problem tackled in the paper can be also viewed as a lattice attached to a single lead (the waveguide) 
which is created by breaking bonds in a pair of semi-infinite rows.
The analysis of the energy flux relative to the waveguide, thus lying between the semi-infinite defects, is 
a derived entity based on the exact solution presented in this paper (schematically shown in Fig. \ref{sq_strip_heatbath_phonons}); the relevant analysis and details shall appear elsewhere.
Additionally, the non-zero offset case 
remains a difficult issue even in the discrete case \cite{GMthesis}, this is not analyzed in this paper.

\section{Lattice model}
\label{latticemodel}
An infinite square lattice, denoted by ${{\mathfrak{S}}}$, as a mechanical structure undergoing anti-plane shear motion, is considered. The lattice consists of identical particles, with in-plane spacing $\la$, each having unit mass and interacting with only its four nearest neighbours through bonds with a spring constant ${1}/{\la^2}$ (see \cite{Bls0} for peculiar choice of scales).
A time harmonic lattice wave is assumed to be incident 
on a pair of rigid constraints or cracks. 
A crack is modeled by assuming that the spring constant 
between the 
particles surrounding the crack is zero, while the rigid constraint is characterized by the vanishing of total displacement at each constrained site.
For convenience, in this paper, sometimes a subscript `${{k}}$' and `${{c}}$' is used to represent an entity associated with case of cracks and rigid constraints, respectively.

Let ${{\Sigma}}_{{k}}$ (resp. ${{\Sigma}}_{{c}}$) denote the set of all lattice sites in ${{\mathfrak{S}}}$ associated with the crack-faces (resp. rigid constraints), {that 
is precisely those sites which miss} one nearest neighbor bond (resp. {those sites whose displacement is restricted to be zero}). {Suppose ${N}$ is a positive integer (greater than $1$).
Let ${\mathbb{Z}}$ denote the set of all integers. 
Let ${{\mathbb{Z}}^2}$ denote the set ${\mathbb{Z}}\times{\mathbb{Z}}$. }
Corresponding to a separation of $2{N}$ (resp. $2{N}-1$), i.e., for even (resp. odd) width ${N_w}$ of waveguide formed between both cracks,
\begin{subequations}
\begin{eqn}
{{\Sigma}}_{{k}}{:=}\{{({\mathtt{x}}, {\mathtt{y}})\in{{{\mathbb{Z}}^2}}: }
{\mathtt{x}}\ge0, {\mathtt{y}}=-{N}, -{N}-1\}\cup\{{({\mathtt{x}}, {\mathtt{y}})\in{{{\mathbb{Z}}^2}}: }
{\mathtt{x}}\ge0, {\mathtt{y}}={N}, {N}-1\},\\
\text{(resp. }{{\Sigma}}'_{{k}}{:=}\{{({\mathtt{x}}, {\mathtt{y}})\in{{{\mathbb{Z}}^2}}: }
{\mathtt{x}}\ge0, {\mathtt{y}}=-{N}+1,-{N}\}\cup\{{({\mathtt{x}}, {\mathtt{y}})\in{{{\mathbb{Z}}^2}}: }
{\mathtt{x}}\ge0, {\mathtt{y}}={N}, {N}-1\}).
\label{defectK}
\end{eqn}
With separation $2{N}$ (resp. $2{N}-1$), i.e., even (resp. odd) ${N_w}$ waveguide within rigid constraints,
\begin{eqn}
{{\Sigma}}_{{c}}&{:=}\{{({\mathtt{x}}, {\mathtt{y}})\in{{{\mathbb{Z}}^2}}: }
({\mathtt{x}}, -{N}-1)\in{{{\mathbb{Z}}^2}}: {\mathtt{x}}\ge0\}\cup\{{({\mathtt{x}}, {\mathtt{y}})\in{{{\mathbb{Z}}^2}}: }
({\mathtt{x}}, +{N})\in{{{\mathbb{Z}}^2}}: {\mathtt{x}}\ge0\},\\
\text{(resp. }{{\Sigma}}'_{{c}}&{:=}\{{({\mathtt{x}}, {\mathtt{y}})\in{{{\mathbb{Z}}^2}}: }
({\mathtt{x}}, -{N})\in{{{\mathbb{Z}}^2}}: {\mathtt{x}}\ge0\}\cup\{{({\mathtt{x}}, {\mathtt{y}})\in{{{\mathbb{Z}}^2}}: }
({\mathtt{x}}, +{N})\in{{{\mathbb{Z}}^2}}: {\mathtt{x}}\ge0\}).
\label{defectC}
\end{eqn}
\label{defectKC}
\end{subequations}
For convenience, 
the two different parities of the width ${N_w}$ 
are represented by a parity bit ${{\,\gimel\,}}$ (${{\,\gimel\,}}=1$ corresponding to the odd case and ${{\,\gimel\,}}=0$ for even). 
The sets ${{\Sigma}}_{{k}}, {{\Sigma}}'_{{k}}, {{\Sigma}}_{{c}}, {{\Sigma}}'_{{c}},$ from \eqref{defectK} and \eqref{defectC}, can be 
alternatively described as the broken bonds exist between ${\mathtt{y}}={N}, {N}-1$ and ${\mathtt{y}}=-{N}+{{\,\gimel\,}}, -{N}-1+{{\,\gimel\,}}$, while the rigid constraints are located at ${\mathtt{y}}={N}$ and ${\mathtt{y}}=-{N}-1+{{\,\gimel\,}}$.
With ${\Sigma}$ representing either of ${{\Sigma}}_{{k}}$, ${{\Sigma}}'_{{k}}$, ${{\Sigma}}_{{c}}$, ${{\Sigma}}'_{{c}}$, the equation of motion 
at $({\mathtt{x}}, {\mathtt{y}})\in{{{\mathbb{Z}}^2}}\setminus{\Sigma}$ is
\begin{eqn}
\od{^2}{t^2}{\su}_{{\mathtt{x}}, {\mathtt{y}}}=\frac{1}{\la^2}{\triangle}{\su}_{{\mathtt{x}}, {\mathtt{y}}}, 
\text{where }
{\triangle}{\su}_{{\mathtt{x}}, {\mathtt{y}}}{:=}{\su}_{{\mathtt{x}}+1, {\mathtt{y}}}+{\su}_{{\mathtt{x}}-1, {\mathtt{y}}}+{\su}_{{\mathtt{x}}, {\mathtt{y}}+1}+{\su}_{{\mathtt{x}}, {\mathtt{y}}-1}-4{\su}_{{\mathtt{x}}, {\mathtt{y}}}.
\label{dimnewtoneq2}
\end{eqn}
\begin{remark}
{
The equation of motion on sites located on the upper and lower face of a crack, respectively, is 
\beqans
\od{^2}{t^2}{\su}_{{\mathtt{x}}, {\mathtt{y}}}=\frac{1}{\la^2}({\su}_{{\mathtt{x}}+1, {\mathtt{y}}}+{\su}_{{\mathtt{x}}-1, {\mathtt{y}}}+{\su}_{{\mathtt{x}}, {\mathtt{y}}+1}-3{\su}_{{\mathtt{x}}, {\mathtt{y}}}), 
\label{dimnewtoneqcracku}\\
\od{^2}{t^2}{\su}_{{\mathtt{x}}, {\mathtt{y}}}=\frac{1}{\la^2}({\su}_{{\mathtt{x}}+1, {\mathtt{y}}}+{\su}_{{\mathtt{x}}-1, {\mathtt{y}}}+{\su}_{{\mathtt{x}}, {\mathtt{y}}-1}-3{\su}_{{\mathtt{x}}, {\mathtt{y}}}).
\label{dimnewtoneqcrackl}
\eeqans{dimnewtoneqcrackall}
}
\label{crackbc}
\end{remark}
\begin{remark}
{
The equation of motion on sites immediately above and below a rigid constraint, respectively, is
\begin{subequations}
\beqan
\od{^2}{t^2}{\su}_{{\mathtt{x}}, {\mathtt{y}}}=\frac{1}{\la^2}({\su}_{{\mathtt{x}}+1, {\mathtt{y}}}+{\su}_{{\mathtt{x}}-1, {\mathtt{y}}}+{\su}_{{\mathtt{x}}, {\mathtt{y}}+1}-4{\su}_{{\mathtt{x}}, {\mathtt{y}}}), 
\label{dimnewtoneqconstraintu}\\
\od{^2}{t^2}{\su}_{{\mathtt{x}}, {\mathtt{y}}}=\frac{1}{\la^2}({\su}_{{\mathtt{x}}+1, {\mathtt{y}}}+{\su}_{{\mathtt{x}}-1, {\mathtt{y}}}+{\su}_{{\mathtt{x}}, {\mathtt{y}}-1}-4{\su}_{{\mathtt{x}}, {\mathtt{y}}}).
\label{dimnewtoneqconstraint}
\eeqan
The equation of motion for the single site facing a semi-infinite rigid constraint is
\begin{eqn}
\od{^2}{t^2}{\su}_{{\mathtt{x}}, {\mathtt{y}}}=\frac{1}{\la^2}({\su}_{{\mathtt{x}}+1, {\mathtt{y}}}+{\su}_{{\mathtt{x}}, {\mathtt{y}}+1}+{\su}_{{\mathtt{x}}, {\mathtt{y}}-1}-4{\su}_{{\mathtt{x}}, {\mathtt{y}}}), 
\label{dimnewtoneqconstraintahead}
\end{eqn}
Indeed, for the sites on each rigid constraint ${\su}_{{\mathtt{x}}, {\mathtt{y}}}=0$.
\label{dimnewtoneqconstraintall}
\end{subequations}
}
\label{clampedbc}
\end{remark}

In this paper, {the considered structure admits two distinct kinds of incident waves: one type of incident wave is the {\em bulk lattice wave} that corresponds to the passband of the square lattice outside the waveguide formed by the two semi-infinite defects, while the second type of incident wave is the {\em lattice waveguide mode} that corresponds to the passband of the waveguide formed by the two semi-infinite defects.} The role of {\em type} of incidence is emphasized by writing `incidence from the bulk lattice' vis-a-vis `{\em incidence from the waveguide}'.
Consider the former and let ${\su}^{\inc B}$ describe the {incident wave} 
with frequency $\icf$ and a {lattice wave vector} $({\upkappa}_x, {\upkappa}_y)$;
specifically, 
\begin{eqn}
{\su}_{{\mathtt{x}}, {\mathtt{y}}}^{\inc B}{:=}{{\mathrm{A}}}e^{i{\upkappa}_x {\mathtt{x}}+i{\upkappa}_y {\mathtt{y}}-i\icf t},
\label{uincB}
\end{eqn}
where ${{\mathrm{A}}}\in{\mathbb{C}}$ is constant {(${\mathbb{C}}$ denotes the set of complex numbers; 
${z}\in{\mathbb{C}},$ ${z}={z}_1+i{z}_2, {z}_1\in{\mathbb{R}}, {z}_2\in{\mathbb{R}}$ with ${\mathbb{R}}$ as the set of real numbers). }
{Following a traditional choice in diffraction theory \cite{Bouwkamp, Noble}, as a way to avoid the technical issues associated with nondecaying wavefronts, a vanishingly small amount of damping is introduced in the lattice model. This leads to a complex $\icf$ with a vanishingly small but positive imaginary part.}
\begin{figure}[t]
\centering
{\includegraphics[width=\textwidth]{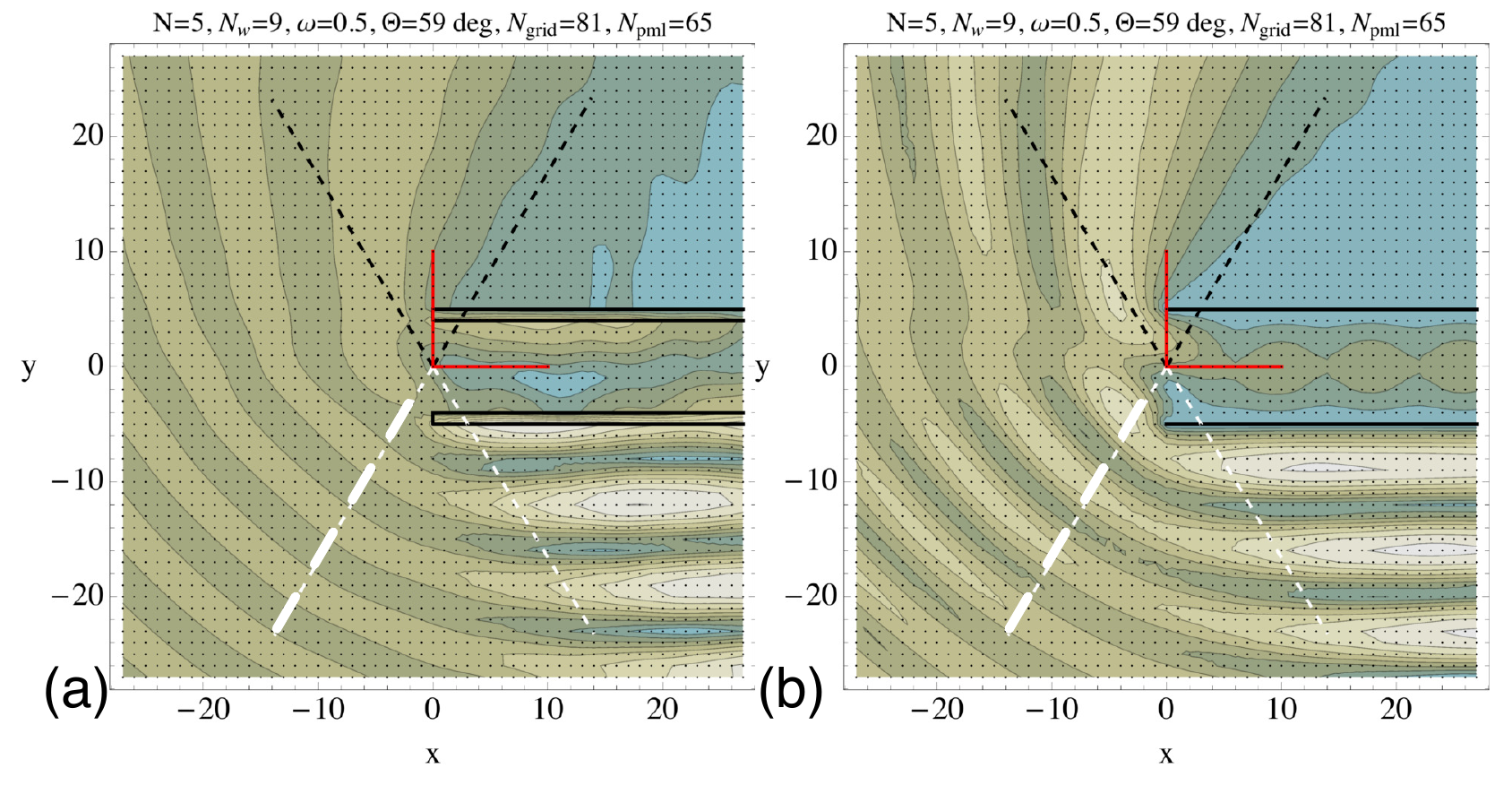}}
\caption{Square lattice with a pair of semi-infinite (a) cracks and (b) rigid constraints. Illustration provides the contourplot of the total wavefield, obtained by numerical scheme (summarized in an Appendix of \cite{Bls0}), in the presence of incident wave \eqref{uincB} (shown as thick white ray). Here ${\,\gimel\,}=1$ as ${N_w}$ is odd.}
\label{utot_parity_0_w_5_Theta_59_Nt_5_defecttype_12}
\end{figure}
Throughout the paper, the 
factor, $e^{-i\icf t}$, is suppressed.
{In the absence of damping,
by} virtue of \eqref{dimnewtoneq2} in intact lattice (${\su}={\su}^{\inc B}$), the triplet ${\upomega}$ (${:=} \la\icf$), ${\upkappa}_x,$ and ${\upkappa}_y$ satisfies the 
dispersion relation 
\begin{eqn}
{\upomega}^2
=4(\sin^2{\tfrac{1}{2}}{\upkappa}_x+\sin^2{\tfrac{1}{2}}{\upkappa}_y), \quad\quad({\upkappa}_x, {\upkappa}_y)\in [-\pi, \pi]^2,
\label{dispersion}
\end{eqn}
{while} the lattice wave \eqref{uincB} is diffracted by the pair of semi-infinite defects as illustrated by Fig. \ref{utot_parity_0_w_5_Theta_59_Nt_5_defecttype_12}.
{
With ${\upomega}={\upomega}_1+i{\upomega}_2, {\upomega}_2>0$, it is easy to see that the {wave number} of the bulk incident lattice wave ${\su}^{\inc B}$ \eqref{uincB} is also a complex number, i.e.,
${\upkappa}={\upkappa}_1+i{\upkappa}_2, 
{\upkappa}_2>0$,
which is related to the complex ${\upkappa}_x$ and ${\upkappa}_y$ through \eqref{dispersion} and the {angle of incidence} ${\Theta}\in(-\pi, \pi]$ of ${\su}^{\inc B}$ so that 
${\upkappa}_x={\upkappa}\cos{\Theta}, {\upkappa}_y={\upkappa}\sin{\Theta}.$ Due to symmetry it is enough to consider ${\Theta}\in(0, \pi]$.
For the assumed model, when ${\upomega}_1\in(2, 2\sqrt{2})$ the allowed values ${\Theta}$ lie in a subset of $(0, \pi]$.
In general, it is assumed that ${\upomega}_1\in[0, 2\sqrt{2}]\setminus S_e,$ where $
S_e=\{0, 2, 2\sqrt{2}\}
$ \cite{Shaban}. The assumption of complex frequency, analogous to above, holds for the {\em incidence from the waveguide} when a wave mode inside the waveguide formed by the two defects replaces the ansatz \eqref{uincB}.
}

Taking cue from the continuum model \cite{Abrahams0,Noble}, with some effort for the discrete model, it is easy to recognize the presence of a $2\times2$ matrix Wiener--Hopf ({{WH}}) kernel \cite{GMthesis,gmtwocracks}; the details are omitted in this paper \cite{GMthesis}. 
{Intuitively, the $2\times2$ matrix {{WH}} kernel arises as} the two sequences of sources on a pair of semi-infinite rows, induced by the defects interacting with incident wave and scattered wave, cannot be de-coupled from each other in the presence of stagger. 
\begin{remark}
{On the lines of \S3 of \cite{Bls2} and \S7 of \cite{Bls3}, it is stated without proof that {\em given ${\upomega}_2>0$ and ${\upomega}_1\in[0, 2\sqrt{2}]\setminus S_e,$ there exists a unique solution of the scattered wave field in $\ell_2({\mathbb{Z}}^2)$}. The proof (omitted in this paper) utilizes the properties of $2\times2$ matrix {{WH}} kernel analogous to those stated as Lemma 3.1 and Lemma 3.2 in \cite{Bls2} and Lemma 7.1 in \cite{Bls3}.
}
\label{uniquenessremark}
\end{remark}
{However, from the viewpoint of explicit solution, going beyond the existence and uniqueness of the solution in Remark \ref{uniquenessremark}, in the special case of the absence of stagger, due to the alignment of the defect tips ({see Fig. \ref{squarelattice_modeIII_2slits_zero_stagger}}), a reduction from infinite lattice ${{\mathfrak{S}}}$ to lattice half-plane, denoted by ${{\mathfrak{S}}}_{\text{H}}$, can be exploited. This is possible due to the geometric reflection symmetry as explained in the next section. }

\section{Geometric symmetry based reduction}
\label{georeduction}
\begin{figure}[t]
\centering
{\includegraphics[width=.65\textwidth]{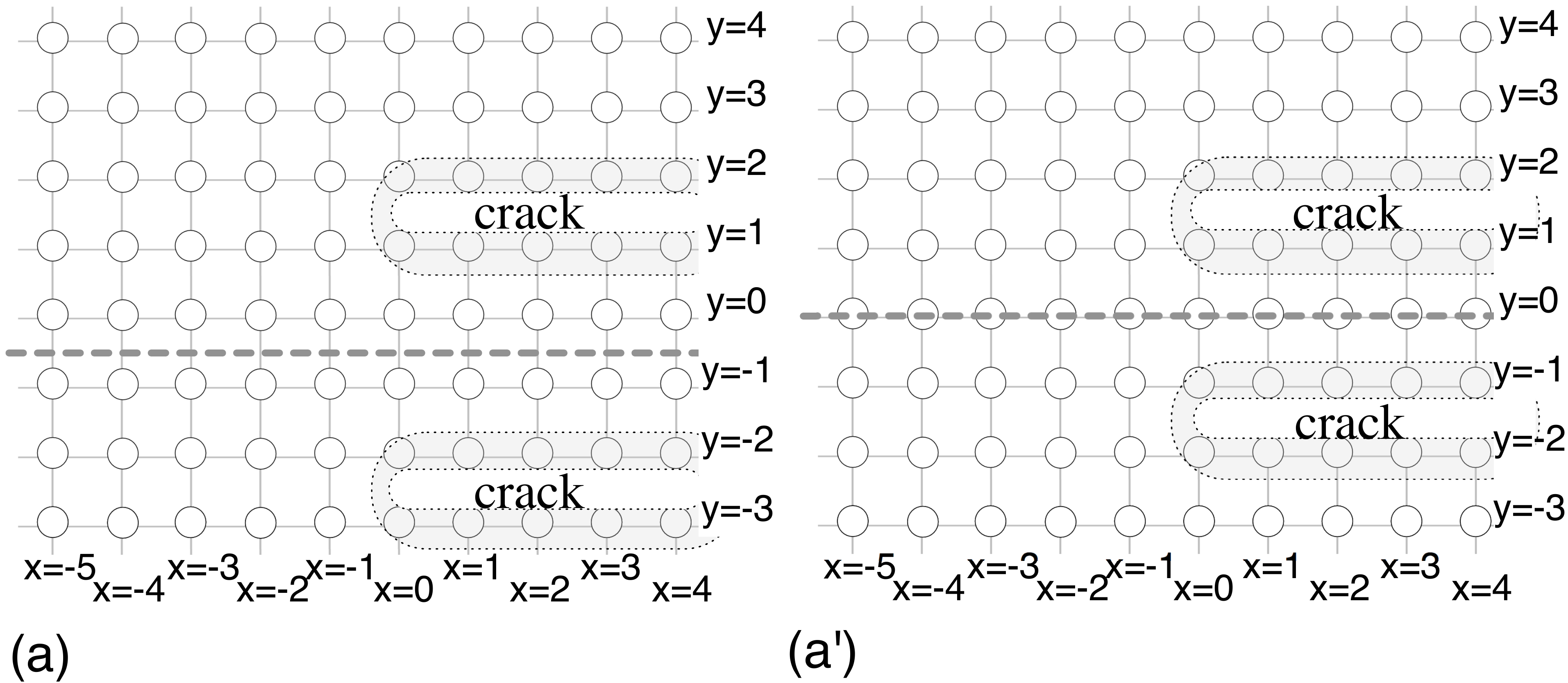}\\}
{\includegraphics[width=.65\textwidth]{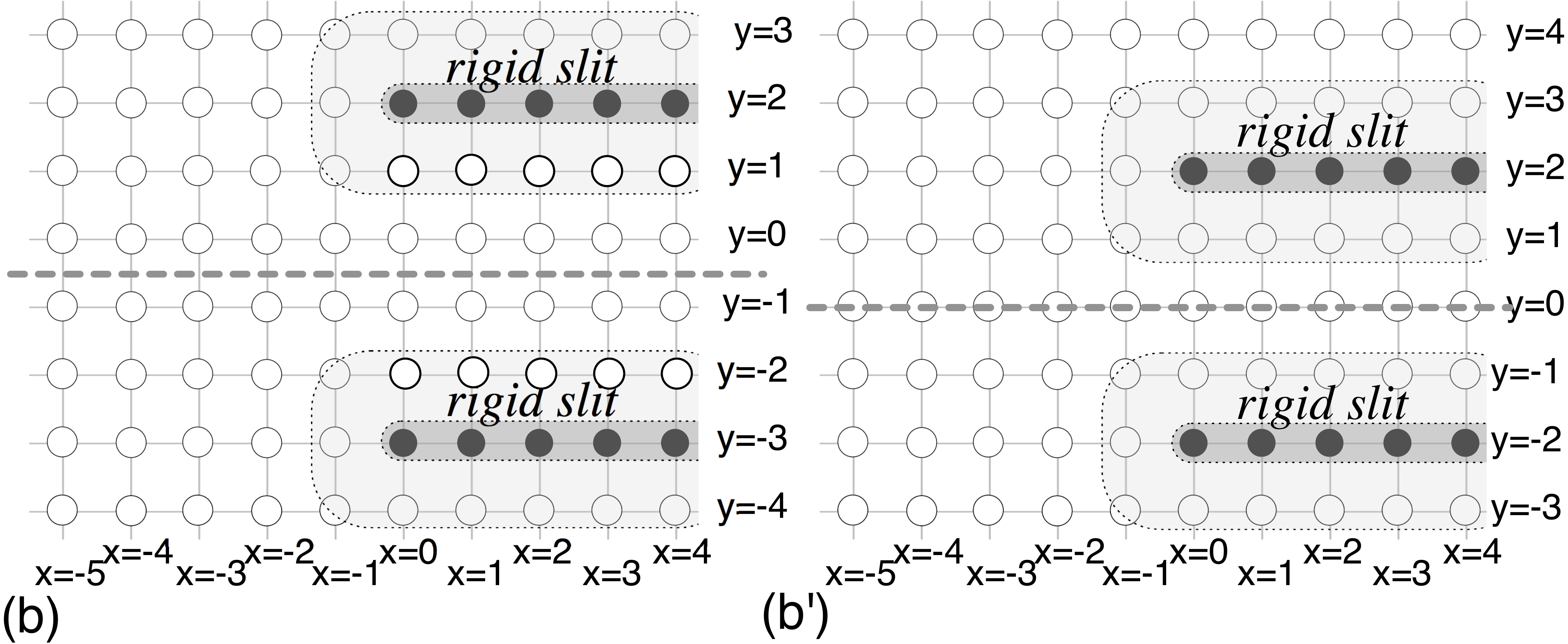}}
\caption{
{
Square lattice with 
the broken vertical bonds between (with ${N}=2$) 
(a) ${\mathtt{y}}=\pm{N}, {\mathtt{y}}=\pm{N}-1,$ for all ${\mathtt{x}}\ge0,$
(a') ${\mathtt{y}}=\pm{N}, {\mathtt{y}}=\pm{N}\mp1$,for all ${\mathtt{x}}\ge0,$
the constrained sites located (with ${N}=2$) at 
(b) ${\mathtt{y}}=\pm{N}-{\tfrac{1}{2}}\pm{\tfrac{1}{2}},$ for all ${\mathtt{x}}\ge0,$ (b') 
${\mathtt{y}}=\pm{N}$, for all ${\mathtt{x}}\ge0.$
}
}
\label{squarelattice_modeIII_2slits_zero_stagger}
\end{figure}
In order to utilize the geometric symmetry in the physical structure, 
it is natural to consider the even/odd symmetry relative the mid-plane {(shown by thick dashed line in Fig. \ref{squarelattice_modeIII_2slits_zero_stagger}).}
According to \eqref{defectKC},
with odd number of rows in-between the defects, 
the waveguide width ${N_w}$ 
is $2{N}-1$,
on the other hand for the 
even number of rows in-between,
the corresponding waveguide width formed by the two rigid constraints and by the two cracks is ${N_w}=2{N}$.
{The main idea behind the reduction to lattice-half plane can be understood as follows}.

\begin{figure}[ht!]
\centering
{\includegraphics[width=\textwidth]{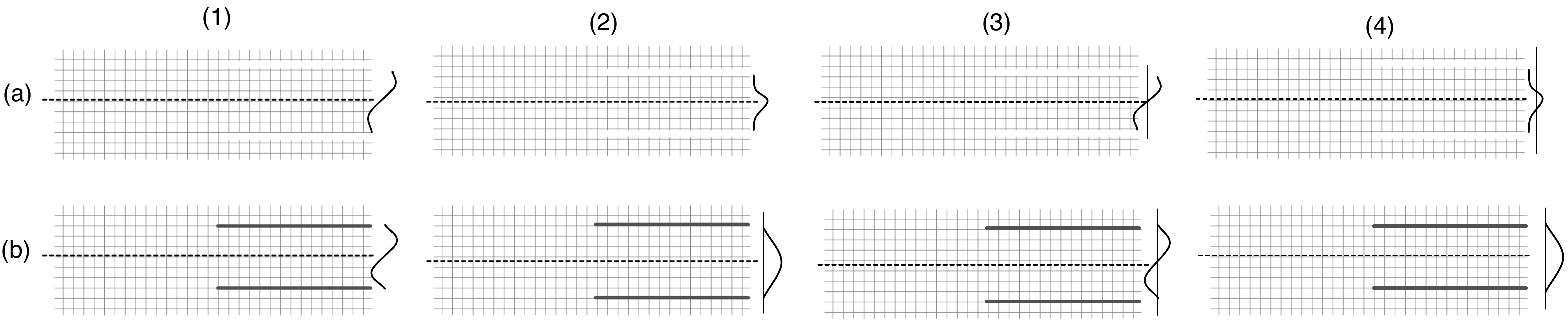}}
\caption{{Square lattice with a pair of semi-infinite cracks or rigid constraints and geometric symmetry related to the incident and outgoing wave. The labels (1)--(4) correspond to the conditons Case H\ref{cond1}--H\ref{cond4}.}}
\label{squarelattice_modeIII_2defects_zero_stagger_geometricsymmetry}
\end{figure}

{Consider the (bulk) incident wave \eqref{uincB}.
Recall Fig. \ref{squarelattice_modeIII_2defects_zero_stagger_geometricsymmetry}.
Two cases arise depending on the even/odd parity of the separation between the two cracks or rigid constraints.
For $({\mathtt{x}}, {\mathtt{y}})\in{{{\mathbb{Z}}^2}}$,
\beqans
{\su}_{{\mathtt{x}}, {\mathtt{y}}}^{\inc B}&=&{\tfrac{1}{2}}({\su}_{{\mathtt{x}}, {\mathtt{y}}}^{\inc B}+{\su}_{{\mathtt{x}}, -{\mathtt{y}}-1}^{\inc B})+{\tfrac{1}{2}}({\su}_{{\mathtt{x}}, {\mathtt{y}}}^{\inc B}-{\su}_{{\mathtt{x}}, -{\mathtt{y}}-1}^{\inc B})\\
&=&{{\mathrm{A}}}e^{i{\upkappa}_x {\mathtt{x}}}e^{-i{\tfrac{1}{2}}{\upkappa}_y}\cos{\upkappa}_y({\mathtt{y}}+{\tfrac{1}{2}})+i{{\mathrm{A}}}e^{i{\upkappa}_x {\mathtt{x}}}e^{-i{\tfrac{1}{2}}{\upkappa}_y}\sin{\upkappa}_y({\mathtt{y}}+{\tfrac{1}{2}}), 
\label{uincBeven}
\eeqans{uincBevenall}
and
\beqans
{\su}_{{\mathtt{x}}, {\mathtt{y}}}^{\inc B}&=&{\tfrac{1}{2}}({\su}_{{\mathtt{x}}, {\mathtt{y}}}^{\inc B}+{\su}_{{\mathtt{x}}, -{\mathtt{y}}}^{\inc B})+{\tfrac{1}{2}}({\su}_{{\mathtt{x}}, {\mathtt{y}}}^{\inc B}-{\su}_{{\mathtt{x}}, -{\mathtt{y}}}^{\inc B})\\
&=&{{\mathrm{A}}}e^{i{\upkappa}_x {\mathtt{x}}}\cos{\upkappa}_y{\mathtt{y}}+i{{\mathrm{A}}}e^{i{\upkappa}_x {\mathtt{x}}}\sin{\upkappa}_y{\mathtt{y}}.
\label{uincBodd}
\eeqans{uincBoddall}
The first term in \eqref{uincBeven} (resp. \eqref{uincBodd}) is even-symmetric relative to ${\mathtt{y}}=-{\tfrac{1}{2}}$ (resp. ${\mathtt{y}}=0$) while the second term is odd-symmetric. Due to the linearity of the scattering problem, using the uniqueness of the solution stated above in Remark \ref{uniquenessremark}, it is clear that the scattered wave field also respects the same symmetry and admits an {\em identical} decomposition where its even-symmetric (resp. odd-symmetric) component corresponds to even-symmetric (resp. odd-symmetric) component of incident wave.
}

For the even symmetry {of the wave field (incident as well as scattered)} in case of even separation, the equivalent reduction to {lattice} half-plane ${{\mathfrak{S}}}_{\text{H}}$ with free boundary condition is, thus, possible {since $\su_{{\mathtt{x}}, {\mathtt{y}}}=\su_{{\mathtt{x}}, -{\mathtt{y}}-1}, {\mathtt{y}}\ge0$ leads to effectively an absence of bond between the rows located at ${\mathtt{y}}=0$ and ${\mathtt{y}}=-1$}. 
Similarly, for the odd symmetry in case of odd separation,
the equivalent reduction to {lattice} half-plane ${{\mathfrak{S}}}_{\text{H}}$ with fixed boundary condition holds {since $\su_{{\mathtt{x}}, {\mathtt{y}}}=-\su_{{\mathtt{x}}, -{\mathtt{y}}}, {\mathtt{y}}\ge0$ leads to a zero displacement condition for the row located at ${\mathtt{y}}=0$}. 
In the other {two} cases the problem becomes equivalent to a {lattice} half-plane problem with {a} slightly different boundary condition; {the details are omitted}. 
See Fig. \ref{squarelattice_modeIII_2defects_zero_stagger_geometricsymmetry} for a {graphical} depiction of the geometric symmetry for the context of a pair of semi-infinite defects.

{The diffraction problems on infinite lattice ${{\mathfrak{S}}}$ involving a pair of semi-infinite defects have been solved in this paper by reduction to a problem on lattice half-plane ${{\mathfrak{S}}}_{\text{H}}$ with a single semi-infinite defect
forming a waveguide with lattice half-plane boundary at ${\mathtt{y}}=0$. For this purpose, consider the following definition
\begin{eqn}
{{\mathbb{Z}}}^2_{\text{H}}{:=}\{({\mathtt{x}}, {\mathtt{y}}): {\mathtt{x}}, {\mathtt{y}}\in{\mathbb{Z}}, {\mathtt{y}}\ge0\}.
\label{halfplane}
\end{eqn}
The coordinates associated with ${{\mathfrak{S}}}_{\text{H}}$, including a single semi-infinite defect, are illustrated in Fig. \ref{squarelattice_modeIII_singledefects_halfplane} {(the same can be contrasted with the choice of coordinates for the infinite lattice as shown in Fig. \ref{squarelattice_modeIII_2slits_zero_stagger})}.}
The condition at half-plane boundary is described by two parameters ${\upbeta}$ and ${\upgamma}$ as
\begin{enumerate}[{Case H}1:]
\item \label{cond1}
${{\upbeta}}=0, {{\upgamma}}=0$ at ${\mathtt{y}}=1$ {for infinite lattice (Fig. \ref{squarelattice_modeIII_2slits_zero_stagger}a', b')}, and at ${\mathtt{y}}=0$ {for the lattice half-plane} the boundary condition {uses ${\su}_{\cdot, {\mathtt{y}}{-1}}=0,$ }
\item \label{cond2}
${{\upbeta}}=0, {{\upgamma}}=-1$ at ${\mathtt{y}}=0$ {for infinite lattice (Fig. \ref{squarelattice_modeIII_2slits_zero_stagger}a, b)} and at ${\mathtt{y}}=0$ {for the lattice half-plane} the boundary condition {uses
${\su}_{\cdot, {\mathtt{y}}-1}=+{\su}_{\cdot, {\mathtt{y}}},$}
\item \label{cond3}
${{\upbeta}}=0, {{\upgamma}}=1$ at ${\mathtt{y}}=0$ {for infinite lattice (Fig. \ref{squarelattice_modeIII_2slits_zero_stagger}a, b)} and at ${\mathtt{y}}=0$ {for the lattice half-plane} the boundary condition {uses ${\su}_{\cdot, {\mathtt{y}}-1}=-{\su}_{\cdot, {\mathtt{y}}},$} 
\item \label{cond4}
${{\upbeta}}=1, {{\upgamma}}=0$ at ${\mathtt{y}}=0$ {for infinite lattice (Fig. \ref{squarelattice_modeIII_2slits_zero_stagger}a', b')} and at ${\mathtt{y}}=0$ {for the lattice half-plane} the boundary condition {uses ${\su}_{\cdot, {\mathtt{y}}-1}={\su}_{\cdot, {\mathtt{y}}+1}$}.
\end{enumerate}
In a general case, that includes Case H\ref{cond1}--H\ref{cond4}, for lattice row at the half-plane boundary {(${\mathtt{y}}=0$)}, 
\begin{eqn}
{\su}_{{\mathtt{x}}+1, {\mathtt{y}}}+{\su}_{{\mathtt{x}}-1, {\mathtt{y}}}+{\su}_{{\mathtt{x}}, {\mathtt{y}}+1}+({\upomega}^2-4){\su}_{{\mathtt{x}}, {\mathtt{y}}}+{{\upbeta}}{\su}_{{\mathtt{x}}, {\mathtt{y}}+1}-{{\upgamma}}{\su}_{{\mathtt{x}}, {\mathtt{y}}}=0.
\label{genBC}
\end{eqn}
Naturally, the scattering occurs due to {a single} semi-infinite defect {along} with an equation of motion \eqref{genBC} (boundary condition) at the edge of {the} half-plane {${{\mathfrak{S}}}_{\text{H}}$} in the presence of the incident wave \eqref{uincB}. 
In view of the reduction ({Fig. \ref{squarelattice_modeIII_2slits_zero_stagger}--}Fig. \ref{squarelattice_modeIII_singledefects_halfplane}),
it is convenient to consider 
a modified expression for the incident wave 
{(derived from \eqref{uincB} using the reduction based on geometric symmetry)} that {itself} satisfies the boundary condition \eqref{genBC};
in particular, 
\beqan
{\su}_{{\mathtt{x}}, {\mathtt{y}}}^{\inc}{:=}{{\mathrm{A}}}e^{i{\upkappa}_x {\mathtt{x}}+i{\upkappa}_y {\mathtt{y}}}+c_B{{\mathrm{A}}}e^{i{\upkappa}_x {\mathtt{x}}-i{\upkappa}_y {\mathtt{y}}}, 
\quad\quad({\mathtt{x}}, {\mathtt{y}})\in{{{\mathbb{Z}}^2}}, {\mathtt{y}}\ge0, 
\label{uinc}
\eeqan
\text{where $c_B$ is given by }
\beqan
c_B
={\mathtt{C}}_B(e^{-i{\upkappa}_y}),\quad
{\mathtt{C}}_B(\zeta){:=} -\frac{\mathcal{F}_B(\zeta)}{\mathcal{F}_B(\zeta^{-1})}, \quad\mathcal{F}_B(\zeta)=\zeta-{{\upbeta}} \zeta^{-1}+{{\upgamma}}.
\label{defCB}
\eeqan
Above expression results {after simplification of} 
$-e^{-i{\upkappa}_y}-c_Be^{i{\upkappa}_y}+{{\upbeta}}(e^{i{\upkappa}_y}+c_Be^{-i{\upkappa}_y})-{{\upgamma}}(1+c_B)=0.$

The general case {of the scattering problem on a lattice half-plane ${{\mathfrak{S}}}_{\text{H}}$ with above boundary condition \eqref{genBC}} involving ${\upbeta}$ and ${\upgamma}$ can be solved using the 
complex analysis as developed in \cite{Bls9s} and \cite{Bls10mixed}. 
Details, {using similar notation}, are provided 
below {while also following the technique introduced in} \cite{Bls0, Bls1}.

\section{Exact solution based on {{WH}} method
}
\label{WHdiscrete}
\begin{figure}[t]
\centering
{\includegraphics[width=\textwidth]{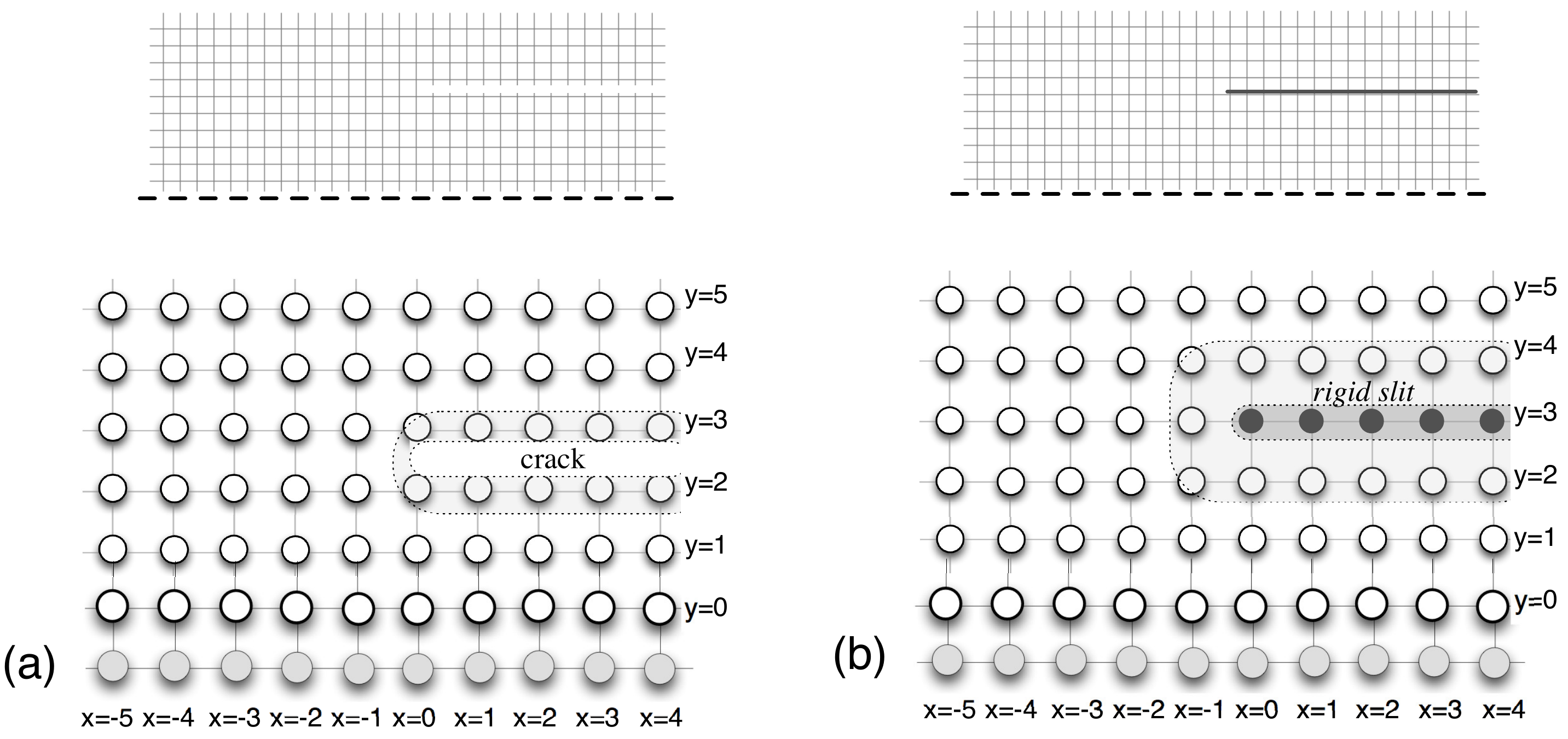}}
\caption{(Top) Reduction based on Fig. \ref{squarelattice_modeIII_2defects_zero_stagger_geometricsymmetry}. (Bottom) Square lattice ${{\mathfrak{S}}}$ with the semi-infinite defect at ${\mathtt{y}}={N}$ ($=3$) and equation of motion \eqref{genBC} at the edge of the half-plane ${{\mathfrak{S}}}_{\text{H}}$ in the presence of the incident wave \eqref{uincB}. 
}
\label{squarelattice_modeIII_singledefects_halfplane}
\end{figure}
{Let (recall \eqref{halfplane})
\beqans
{{\Sigma}}^{\text{H}}_{{k}}&{:=}&\{{({\mathtt{x}}, {\mathtt{y}})\in{{\mathbb{Z}}}^2_{\text{H}}: }
{\mathtt{x}}\ge0, {\mathtt{y}}={N}, {N}-1\},\\
{{\Sigma}}^{\text{H}}_{{c}}&{:=}&\{{({\mathtt{x}}, {\mathtt{y}})\in{{\mathbb{Z}}}^2_{\text{H}}: }
({\mathtt{x}}, {N})\in{{{\mathbb{Z}}^2}}: {\mathtt{x}}\ge0\}.
\eeqans{}
Above sets correspond to the crack and rigid constraint provided in the schematic illustration of Fig. \ref{squarelattice_modeIII_singledefects_halfplane}a, b, respectively.}
The total field ${\su}^{\totwave}$ at an arbitrary site in {${{\mathfrak{S}}}_{\text{H}}$} is a sum of the incident wave field ${\su}^{\inc}$ {\eqref{uinc}} and the scattered field ${\su}^{\scawave}$. 
For simplicity, the letter ${\su}$ is used in place of ${\su}^{\scawave}$. 
By \eqref{dimnewtoneq2} and the definition of ${\upomega}$, the total field ${\su}^{\totwave}$ satisfies the discrete Helmholtz equation
\begin{eqn}
{\triangle}{\su}^{\totwave}_{{\mathtt{x}}, {\mathtt{y}}}+{\upomega}^2{\su}^{\totwave}_{{\mathtt{x}}, {\mathtt{y}}}=0, 
\quad({\mathtt{x}}, {\mathtt{y}})\in{{{\mathbb{Z}}}^2_{\text{H}}\setminus{{\Sigma}}}, 
\label{dHelmholtz}
\text{where }{\su}_{{\mathtt{x}}, {\mathtt{y}}}^{\totwave}={\su}_{{\mathtt{x}}, {\mathtt{y}}}^{\inc}+{\su}_{{\mathtt{x}}, {\mathtt{y}}}, ({\mathtt{x}}, {\mathtt{y}})\in{{{\mathbb{Z}}^2}},
\end{eqn}
except on {the single} rigid constraint ${{\Sigma}}{={\Sigma}^{\text{H}}_{{c}}}$ {(the equation corresponding to \eqref{dimnewtoneqconstraintall} holds in the sites near the constraint)} or {the crack-faces of the single crack} ${{\Sigma}}{={\Sigma}^{\text{H}}_{{k}}}$ {(where the equation corresponding to \eqref{dimnewtoneqcrackall} holds)}, while at (half-plane boundary)
${\mathtt{y}}=0$ \eqref{genBC} holds.

{
Let the letter ${{\mathcal{H}}}$ stands for the Heaviside function: ${{\mathcal{H}}}({{\mathtt{x}}})=0, {{\mathtt{x}}}<0$ and ${{\mathcal{H}}}({{\mathtt{x}}})=1, {{\mathtt{x}}\ge0}$. 
The discrete Fourier transform \cite{Bls0} of the scattered field $\{{\su}_{{\mathtt{x}}, {\mathtt{y}}}\}_{{\mathtt{x}}\in{\mathbb{Z}}}$ at given ${\mathtt{y}}\in{\mathbb{Z}}$ is defined by 
\begin{eqn}
{\su}_{{\mathtt{y}}}^F{:=}{\su}_{{\mathtt{y}}; +}+{\su}_{{\mathtt{y}}; -}, {\su}_{{\mathtt{y}}; \pm}=\sum\nolimits_{{{\mathtt{x}}}=-\infty}^{+\infty}{{z}}^{-{{\mathtt{x}}}}{{\mathcal{H}}}(\pm{{\mathtt{x}}}-{\tfrac{1}{2}}\pm{\tfrac{1}{2}}){\su}_{{{\mathtt{x}}}, {\mathtt{y}}}.
\label{discreteFT}\end{eqn}
In this paper, 
${{z}}$ 
denotes the complex variable after the application of 
Fourier transform. }
By an application of (discrete) 
Fourier transform {\eqref{discreteFT}} (see {also other details in} Appendix \ref{wellposedness}), 
in view of the form of incident wave \eqref{uinc} and splitting of the total wave field, the condition \eqref{genBC} 
at ${\mathtt{y}}=0$ becomes
\beqan
(\sQ+{{\upgamma}}){\su}^F_{0}=(1+{{\upbeta}}){\su}^F_{1},
\label{genBC_FT}
\quad\quad
\text{where }
{\sQ}({{z}}){:=}4-{{z}}-{{z}}^{-1}-{\upomega}^2, {{z}}\in{\mathbb{C}}.
\eeqan

\subsection{Crack}
\label{WHdiscretecrack}
{Let ${\mathbb{Z}}_a^b$ denote the set of integers $\{a, a+1, \dotsc, b\}$.}
Using the definition of $\lambda$ \eqref{lambdadef} \cite{Bls0,Bls1,Slepyanbook}, the {Fourier transform of the (scattered component of the)} solution of eq. \eqref{dHelmholtz} is expressed as
\begin{eqn}
{\su}_{{\mathtt{y}}}^F=\su^F_{{N}}{{\lambda}}^{{\mathtt{y}}-{{N}}}, 
\quad\quad
{\su}_{{\mathtt{y}}}^F
={\su}^F_0(\frac{{{\lambda}}^{-2{{N}}+2}{{\lambda}}^{{\mathtt{y}}}-{{\lambda}}^{-{\mathtt{y}}}}{{{\lambda}}^{-2{{N}}+2}-1})+{\su}^F_{{{N}}-1}(\frac{{{\lambda}}^{-{{N}}+1}{{\lambda}}^{-{\mathtt{y}}}-{{\lambda}}^{-{{N}}+1}{{\lambda}}^{{\mathtt{y}}}}{{{\lambda}}^{-2{{N}}+2}-1}).
\label{ubulkK}
\end{eqn}
for ${{\mathtt{y}}} \ge {{N}}$ and ${\mathtt{y}}\in{\mathbb{Z}}_0^{{N}-1}$, respectively.
Note that
\begin{eqn}
{\su}_{1}^F=f_1{\su}^F_0+g_1{\su}^F_{{{N}}-1},\quad {\su}_{{{N}}-2}^F=f_{{{N}}-2}{\su}^F_0+g_{{{N}}-2}{\su}^F_{{{N}}-1},\quad {\su}_{{{N}}+1}^F=\su^F_{{N}}{{\lambda}},
\label{ubulkKspec}
\end{eqn}
\begin{eqn}
\text{where }
f_1&
{:=}\frac{{{\lambda}}^{-{{N}}+2}-{{\lambda}}^{{{N}}-2}}{{{\lambda}}^{-2{{N}}+2}-1}{{\lambda}}^{-{{N}}+1}, 
f_{{{N}}-2}
{:=}\frac{{{\lambda}}^{-1}-{{\lambda}}}{{{\lambda}}^{-2{{N}}+2}-1}{{\lambda}}^{-{{N}}+1}, 
g_1
=f_{{{N}}-2}, 
g_{{{N}}-2}
=f_1.
\label{fgK}
\end{eqn}
By \eqref{genBC_FT} and \eqref{ubulkKspec},
${\su}_{0}^F$ can be expressed in terms of ${\su}_{{{N}}-1}^F,$
\begin{eqn}
{\su}_{0}^F=\frac{(1+{{\upbeta}})f_{{{N}}-2}}{(\sQ+{{\upgamma}}-(1+{{\upbeta}})f_1)}{\su}_{{{N}}-1}^F, 
\label{u0uNn1K}
\end{eqn}
thereby reducing the set of unknown functions in \eqref{ubulkK} to ${\su}_{{{N}}}^F$ and ${\su}_{{{N}}-1}^F$.
For the lattice row at ${\mathtt{y}}={{N}}-1$ and ${\mathtt{y}}={{N}}, $ respectively,
\beqans-{\upomega}^2{\su}_{{\mathtt{x}}, {{N}}-1}+({\su}_{{\mathtt{x}}, {{N}}-1}-{\su}_{{\mathtt{x}}, {{N}}}){\mathcal{H}}(-{\mathtt{x}}-1)=({\su}^{\inc}_{{\mathtt{x}}, {{N}}-1}-{\su}^{\inc}_{{\mathtt{x}}, {{N}}}){\mathcal{H}}({\mathtt{x}})+{\su}_{{\mathtt{x}}+1, {{N}}-1}\notag\\+{\su}_{{\mathtt{x}}-1, {{N}}-1}+{\su}_{{\mathtt{x}}, {{N}}-2}-3{\su}_{{\mathtt{x}}, {{N}}-1},\\-{\upomega}^2{\su}_{{\mathtt{x}}, {{N}}}+({\su}_{{\mathtt{x}}, {{N}}}-{\su}_{{\mathtt{x}}, {{N}}-1}){\mathcal{H}}(-{\mathtt{x}}-1)=({\su}^{\inc}_{{\mathtt{x}}, {{N}}}-{\su}^{\inc}_{{\mathtt{x}}, {{N}}-1}){\mathcal{H}}({\mathtt{x}})+{\su}_{{\mathtt{x}}+1, {{N}}}\notag\\+{\su}_{{\mathtt{x}}{{N}}-1, {{N}}}+{\su}_{{\mathtt{x}}, {{N}}+1}-3{\su}_{{\mathtt{x}}, {{N}}}.\eeqans{twocracksallN}
{Following commonly used notation, It is supposed that $|{z}|$ denotes the modulus and $\arg {z}$ denotes the argument (with branch cut along negative real axis) for ${z}\in{\mathbb{C}}$. }
\text{Let }
\beqan
\delta_{D+}({z}){:=}\sum\nolimits_{{\mathtt{x}}=0}^\infty{z}^{-{\mathtt{x}}}=(1-{z}^{-1})^{-1},~|{z}|>1,
\label{delDp}
\text{and }
{{z}}_{{P}}{:=} e^{i{\upkappa}_x},\\
{\sv}^{\inc}_{{N}; +}{:=}\sum\nolimits_{{\mathtt{x}}=0}^\infty{{z}}^{-{\mathtt{x}}}({\su}^{\inc}_{{\mathtt{x}}, {{N}}}-{\su}^{\inc}_{{\mathtt{x}}, {{N}}-1})
={\sv}^{\inc}_{0,{N}}\delta_{D+}({{z}} {z}_{{P}}^{-1}),
\label{svinc}\\
\text{with }
{\sv}^{\inc}_{0,{N}}{:=}{{\mathrm{A}}}(e^{i{\upkappa}_y{N}}+c_Be^{-i{\upkappa}_y{N}}-e^{-i{\upkappa}_y}e^{i{\upkappa}_y{N}}-c_Be^{i{\upkappa}_y}e^{-i{\upkappa}_y{N}}).
\label{svinc0N}
\eeqan
{As special case of \eqref{discreteFT}}, using the definitions 
${\su}_{{{N}}; \pm}=\sum\nolimits_{{{\mathtt{x}}}=-\infty}^{+\infty}{{z}}^{-{{\mathtt{x}}}}{{\mathcal{H}}}(\pm{{\mathtt{x}}}-{\tfrac{1}{2}}\pm{\tfrac{1}{2}}){\su}_{{{\mathtt{x}}}, {N}},\quad {\su}_{{N}}^F=({\su}_{{{N}}; +}+{\su}_{{{N}}; -}).$
and similar definitions for 
for ${\su}_{{N}+1; \pm}$ and ${\su}_{{{N}}-1; \pm},$ taking the Fourier transform of \eqref{twocracksallN}, using \eqref{ubulkKspec}
and \eqref{defh2} {(i.e. $\sH=\sQ-2$)}, it is found that
\beqans
({\sH}+1-{f_1})({\su}_{{{N}}-1; +}+{\su}_{{{N}}-1; -})+({\su}_{{{N}}-1; -}-{\su}_{{{N}}; -})&=&-{\sv}^{\inc}_{{N}; +}+{\su}_{0}^F{f_{{{N}}-2}}, \label{twocracksNn1}\\
({\sH}+1-{{\lambda}})({\su}_{{{N}}; +}+{\su}_{{{N}}; -})-({\su}_{{{N}}-1; -}-{\su}_{{{N}}; -})&=&{\sv}^{\inc}_{{N}; +}.\label{twocracksN}
\eeqans{twocracksall}
Using \eqref{twocracksN},
${\su}_{{{N}}-1; -}=
({{\lambda}}^{-1}-1)({\su}_{{{N}}; +}+{\su}_{{{N}}; -})+{\su}_{{{N}}; -}-{\sv}^{\inc}_{{N}; +}$
after substitution in \eqref{twocracksNn1}, 
\begin{eqn}
({{\lambda}}^{-1}-1)({\su}_{{{N}}; +}+{\su}_{{{N}}; -})&=
{f_{{{N}}-2}}{\su}_{0}^F-({\sH}+1-{f_1})({\su}_{{{N}}-1; +}+{\su}_{{{N}}-1; -}).\label{uN}
\end{eqn}
Let the 
vertical bondlengths in cracked row {(i.e., between ${\mathtt{y}}={N}$ and ${\mathtt{y}}={N}-1$)} be defined by
\begin{eqn}
{\sv}_{{\mathtt{x}}, {{N}}}{:=}{\su}_{{\mathtt{x}}, {{N}}}-{\su}_{{\mathtt{x}}, {{N}}-1}.
\label{vdef}
\end{eqn}
With
${\sv}_{{N}}^F={\sv}_{{{N}}; +}+{\sv}_{{{N}}; -},$
using \eqref{vdef} 
in \eqref{uN}, 
\begin{eqn}
({{\lambda}}^{-1}-1)({\sv}_{{{N}}; +}+{\sv}_{{{N}}; -})
&=
{f_{{{N}}-2}}{\su}_{0}^F-({\sH}-{f_1}+{{\lambda}}^{-1})({\su}_{{{N}}-1; +}+{\su}_{{{N}}-1; -}),
\label{vNnrel}
\end{eqn}
which, using \eqref{u0uNn1K}, is 
an algebraic equation (yielding 
${\su}_{{{N}}-1}^{F}$ in terms of ${\sv}_{{{N}}}^{F}$).
In fact, 
\begin{eqn}
{\sv}_{{N}}^F
&={\mathtt{V}}_{{k}}{\su}_{{{N}}-1}^F, 
{\mathtt{V}}_{{k}}={f_{{{N}}-2}}\frac{(1+{{\upbeta}})f_{{{N}}-2}}{({{\lambda}}^{-1}-1)(\sQ+{{\upgamma}}-(1+{{\upbeta}})f_1)}-\frac{({\sH}-{f_1}+{{\lambda}}^{-1})}{({{\lambda}}^{-1}-1)}.
\label{LamK}
\end{eqn}
By \eqref{vdef}, ${\su}_{{{N}}; +}+{\su}_{{{N}}; -}={\su}_{{{N}}-1; +}+{\su}_{{{N}}-1; -}+{\sv}_{{{N}}; +}+{\sv}_{{{N}}; -}.$
Using {the Equation} \eqref{twocracksN} and {that} ${\sQ}=\sH+2={\lambda}+{\lambda}^{-1}$, {it is found that}
\begin{eqn}
{\sv}_{{{N}}-}={\su}_{{{N}}; -}-{\su}_{{{N}}-1; -}=
-({{\lambda}}^{-1}-1)({\su}_{{{N}}; +}+{\su}_{{{N}}; -})+{\sv}^{\inc}_{{N}; +}.\label{vNn}
\end{eqn}
Finally, the {{WH}} equation obtained for ${\sv}_{{N}}$ is
\beqan
{\mathtt{L}}{\sv}_{{{N}}; +}+{\sv}_{{{N}}; -}=(1-{\mathtt{L}}){\sv}^{\inc}_{{N}; +},
\label{WHK}
\text{ where }
{\mathtt{L}}
=\frac{1+{\mathtt{V}}_{{k}}}{1+(1-{{\lambda}})^{-1}{\mathtt{V}}_{{k}}}
=
{\mathfrak{F}}_{{k}}{{\mathtt{L}}}_{{k}},\\
\text{with the structure factor }
{\mathfrak{F}}_{{k}}({z}; {{\upbeta}}, {{\upgamma}}, {{N}})=1-{\mathtt{C}}_B({\lambda}){{\lambda}}^{2 {N}-1},
\label{defFcrack}
\eeqan
using the definition of ${\mathtt{C}}_B$ given by \eqref{defCB} and ${{\mathtt{L}}}_{{k}}$ given by \cite{Bls0} (i.e., ${{{\mathtt{L}}}_{{k}}=}{\sh}/{\sr}$, see \eqref{lambdadef} for {the details concerning} $\sh$ and $\sr$).
{The {{WH}} equation \eqref{WHK} is posed on an annulus
${\mathscr{A}}$ in the complex plane; the definition is provided in \eqref{annAAL}.}

{
\begin{remark}
Note that as ${N}\to\infty$, the strip lemma holds \cite{Bls9s}, that is, the reduced half plane problem coincides with that due to a single crack on an infinite square lattice \cite{Bls0} in this limit. Consider a disk $B_R$ of {\em fixed} radius $R>1$ centered at the crack tip. Then for $({\mathtt{x}}, {\mathtt{y}})\in{{\mathbb{Z}}^2}\cap B_R$, it is stated without proof that the scattered displacement field $\su_{{\mathtt{x}}, {\mathtt{y}}}$ converges in $\ell_2({{\mathbb{Z}}^2}\cap B_R)$ to that corresponding to a single crack as ${N}\to\infty$. The proof utilizes the assumption that ${\upomega}_2>0$ and properties of kernel same as those stated as Lemma 3.1 and Lemma 3.2 in \cite{Bls2}.  
\label{limitN}
\end{remark}
}

According to \eqref{defCB} and \eqref{svinc0N}, with the notation ${\lambda}_{{P}}{:=}{\lambda}({z}_{{P}})$,
\begin{eqn}
{\sv}^{\inc}_{0,{N}}
&={{\mathrm{A}}}({\lambda}_{{P}}^{{N}}-{\lambda}_{{P}}^{{N}-1})(1-{\mathtt{C}}_B({\lambda}_{{P}}^{-1}){\lambda}_{{P}}^{-2{N}+1}).
\label{svinc0N_bulkinc}
\end{eqn}
Above can be {also} re-written as
${\sv}^{\inc}_{0,{N}}
={{\mathrm{A}}}(e^{i{\upkappa}_y {N}}-e^{i({N}-1){\upkappa}_y})(1-c_Be^{-i{\upkappa}_y (2{N}-1)}).$
Using the multiplicative factorization ${{\mathtt{L}}}_{{}}={{\mathtt{L}}}_{{}+}{{\mathtt{L}}}_{{}-}$, the {{WH}} equation \eqref{WHK} becomes 
\begin{eqn}
{\mathtt{L}}_+{\sv}_{{{N}}; +}+{\mathtt{L}}_-^{-1}{\sv}_{{{N}}; -}={\mathtt{C}}, \quad\quad
{\mathtt{C}}=({\mathtt{L}}_-^{-1}-{\mathtt{L}}_+){\sv}^{\inc}_{{N}; +},
\label{WHKn}
\end{eqn}
{on the annulus
${\mathscr{A}}$.}
An additive factorization \cite{Noble} ${{\mathtt{C}}}={{\mathtt{C}}}_{+}+{{\mathtt{C}}}_{-}$ with
\begin{eqn}
{{\mathtt{C}}}_\pm({{z}})&=
\pm{\sv}^{\inc}_{0,{N}}({{\mathtt{L}}}^{-1}_{{}-}({{z}}_{{P}})-{{\mathtt{L}}}_{{}\pm}^{\pm1}({{z}}))
\delta_{D+}({{z}} {{z}}_{{P}}^{-1}), \quad{{z}}\in{\mathscr{A}},
\label{CpmK}
\end{eqn}
and a standard reasoning based on the Liouville's theorem 
leads to the exact solution 
\begin{eqn}
{\sv}_{{N}; \pm}({{z}})&={{\mathtt{C}}}_{\pm}({{z}}){{\mathtt{L}}}_{\pm}^{\mp1}({{z}}),\quad
{{z}}\in{\mathbb{C}}, |{{z}}|\gtrless
\bfrac{\max}{\min}
\{{{\mathrm{R}}}_{\pm}, {{\mathrm{R}}}_{L_{{}}}^{\pm1}\}.
\label{vNpmK}
\end{eqn}
{The definitions of ${{\mathrm{R}}}_{\pm}$ and ${{\mathrm{R}}}_{L}$ are provided in Appendix \ref{wellposedness}.}
The complex function $\sv^F$, for ${{z}}\in{{\mathscr{A}}}$, is found to be 
\begin{eqn}
{\sv}_{{N}}^F({{z}})&={{\mathrm{A}}}{\mathtt{C}}_0\frac{{{z}} {{\mathtt{K}}}({{z}})}{{{z}}-{{z}}_{{P}}}, 
{{\mathtt{K}}}{:=}(1-{{\mathtt{L}}}^{-1}){{\mathtt{L}}}_-, 
{\mathtt{C}}_0{:=}-{{\mathrm{A}}}^{-1}{\sv}^{\inc}_{0,{N}}{{\mathtt{L}}}^{-1}_{{}-}({{z}}_{{P}})
\in{\mathbb{C}}.
\label{vNzsol}
\end{eqn}
Note that (using \eqref{WHK})
$1-{{\mathtt{L}}}^{-1}={{\lambda} {\mathtt{V}}_{{k}}}/({({\lambda}-1) ({\mathtt{V}}_{{k}}+1)}).$
As an example of a closed form solution in the context of near-tip field analysis (on the lines of \cite{Bls2,Bls3}), 
\eqref{vNpmK} gives
${\sv}_{0,{N}}=\lim_{{z}\to\infty}{\sv}_{{N}; +}({{z}})={{\mathtt{C}}}_{+}({\infty}){{\mathtt{L}}}_{+}^{-1}({\infty})$,
i.e., 
${\sv}^{\totwave}_{0,{N}}={\sv}^{\inc}_{0,{N}}{{\mathtt{L}}}_{+}^{-1}({\infty}){{\mathtt{L}}}^{-1}_{{}-}({{z}}_{{P}}).$

In case of {\em incidence from the waveguide}, the scattering occurs due to the intact bonds ahead of the waveguide. In contrast to \eqref{uinc}, the incident wave is given by
\begin{eqn}
{{\su}_{{\mathtt{x}}, {\mathtt{y}}}^{\inc}{:=}{{\mathrm{A}}}{{a}}_{({{{\kappa}}^{\inc}}){{\mathtt{y}}}}e^{i{\upkappa}_x {\mathtt{x}}}, 
\quad\quad
{\mathtt{x}}\in{\mathbb{Z}}, {\mathtt{y}}\in{\mathbb{Z}}_0^{{N}-1}, }
\label{uinc_altinc}
\end{eqn}
where ${{\mathrm{A}}}$ is a constant and ${{a}}_{({{{\kappa}}^{\inc}}){{\mathtt{y}}}}$ refers to the eigenmode representing a propagating wave in the lattice waveguide formed by {the boundary of
lattice half-plane ${{\mathfrak{S}}}_{\text{H}}$} and the lower side of crack. Notice that ${{a}}_{({{{\kappa}}^{\inc}}){{\mathtt{y}}}}$ automatically satisfies the free boundary condition at ${\mathtt{y}}={N}-1.$
For the lattice row at ${\mathtt{y}}={{N}}-1$ and ${\mathtt{y}}={{N}}, $ respectively,
in place of \eqref{twocracksallN},
\beqans-{\upomega}^2{\su}_{{\mathtt{x}}, {{N}}-1}+({\su}_{{\mathtt{x}}, {{N}}-1}-{\su}_{{\mathtt{x}}, {{N}}}){\mathcal{H}}(-{\mathtt{x}}-1)=-({\su}^{\inc}_{{\mathtt{x}}, {{N}}-1}-{\su}^{\inc}_{{\mathtt{x}}, {{N}}}){\mathcal{H}}(-{\mathtt{x}}-1)+{\su}_{{\mathtt{x}}+1, {{N}}-1}\notag\\+{\su}_{{\mathtt{x}}-1, {{N}}-1}+{\su}_{{\mathtt{x}}, {{N}}-2}-3{\su}_{{\mathtt{x}}, {{N}}-1},\\-{\upomega}^2{\su}_{{\mathtt{x}}, {{N}}}+({\su}_{{\mathtt{x}}, {{N}}}-{\su}_{{\mathtt{x}}, {{N}}-1}){\mathcal{H}}(-{\mathtt{x}}-1)=-({\su}^{\inc}_{{\mathtt{x}}, {{N}}}-{\su}^{\inc}_{{\mathtt{x}}, {{N}}-1}){\mathcal{H}}(-{\mathtt{x}}-1)+{\su}_{{\mathtt{x}}+1, {{N}}}\notag\\+{\su}_{{\mathtt{x}}{{N}}-1, {{N}}}+{\su}_{{\mathtt{x}}, {{N}}+1}-3{\su}_{{\mathtt{x}}, {{N}}}.
\eeqans{twocracksallN_altinc}
As before, ${\mathtt{L}}{\sv}_{{{N}}; +}+{\sv}_{{{N}}; -}=-(1-{\mathtt{L}}){\sv}^{\inc}_{{N}; -}$ follows, 
as an analogue of \eqref{WHK},
and the multiplicative factorization ${{\mathtt{L}}}={{\mathtt{L}}}_{+}{{\mathtt{L}}}_{-}$ leads to {{WH}} equation \eqref{WHKn} with the exception that the right hand side is (for ${{z}}\in{{\mathscr{A}}}$)
${{\mathtt{C}}}({{z}})=({{{\mathtt{L}}}_{+}({{z}})}-{{\mathtt{L}}}^{-1}_{-}({{z}})){\sv}^{\inc}_{0,{N}}\delta_{D-}({{z}} {{z}}_{{P}}^{-1}),$
\text{with}
$\delta_{D-}({z})=\sum\nolimits_{{\mathtt{x}}=-1}^{-\infty}{z}^{-{\mathtt{x}}}={z}(1-{z})^{-1},~|{z}|<1.$
\begin{remark}
{The annulus ${\mathscr{A}}$ involves ${{\mathrm{R}}}_{-}=e^{+{\upkappa}_2}$ with ${\upkappa}$ replaced by $-{\upkappa}_x$, while other details remain same as in the case of bulk incidence (see analogous arguments in \cite{Bls9s})}.
\label{annuluswginc}
\end{remark}
An additive factorization, ${{\mathtt{C}}}={{\mathtt{C}}}_{+}+{{\mathtt{C}}}_{-},$ is 
\begin{eqn}
{{\mathtt{C}}}_\pm({{z}})&=\mp{\sv}^{\inc}_{0,{N}}({{{\mathtt{L}}}_{+}({{z}}_{{P}})}-{{\mathtt{L}}}_{\pm}^{\pm1}({{z}}))\delta_{D-}({{z}} {{z}}_{{P}}^{-1}), \quad{{z}}\in{\mathscr{A}}.
\label{CpmK_altinc}
\end{eqn}
Note that ${\sv}^{\inc}_{0,{N}}$ denotes the expression provided in \eqref{svinc0N}. 
It is easy to see that ${{\mathtt{C}}}_+({{z}})$ and ${{\mathtt{C}}}_-({{z}})$ are analytic at ${{z}}\in{\mathbb{C}}$ with $|{{z}}|>\max\{{{\mathrm{R}}}_+, {{\mathrm{R}}}_{L_{}}\},$ $|{{z}}|<\min\{{{\mathrm{R}}}_-, {{\mathrm{R}}}_{L_{}}^{-1}\}$, respectively.
Further,
${\sv}_{0,{N}}=\lim_{{z}\to\infty}{\sv}_{{N}; +}({{z}})$,
i.e., 
${\sv}^{\totwave}_{0,{N}}={\sv}^{\inc}_{0,{N}}{{\mathtt{L}}}_{+}^{-1}({\infty}){{\mathtt{L}}}^{-1}_{{}-}({{z}}_{{P}}).$
By a reasoning based on the Liouville's theorem \cite{Noble}, the discrete {{WH}} equation is solved and in terms of the one-sided Fourier transform {\eqref{discreteFT}}, 
$\sv^F_{{{N}}}$ is given by \eqref{vNpmK} and
\eqref{vNzsol} with
\begin{eqn}
{\mathtt{C}}_0{:=}-{\sv}^{\inc}_{0,{N}}{{\mathtt{L}}}_{+}({{z}}_{{P}})\in{\mathbb{C}},
\label{vNzsol_altinc}
\end{eqn}
in place of \eqref{vNzsol}${}_3$.

By {the} inverse 
Fourier transform, 
\begin{eqn}
\sv_{{\mathtt{x}},{N}}=\frac{1}{2\pi i}\oint_{{\mathcal{C}}} \sv_{{N}; \pm}({{z}}){{z}}^{{\mathtt{x}}-1}d{{z}},\quad\quad {\mathtt{x}}\in{\mathbb{Z}}, {{\mathtt{x}}\bfrac{\ge}{<}0},
\label{inversediscreteFT}
\end{eqn}
where ${{\mathcal{C}}}_{{z}}$ is a rectifiable, closed, counterclockwise contour in the annulus ${{\mathscr{A}}}$ (recall \eqref{annAAL} {and Remark \ref{annuluswginc}}), and upon substitution of \eqref{vNpmK}, \eqref{CpmK} and \eqref{CpmK_altinc}, the exact expression can be constructed. By \eqref{LamK} and \eqref{vNzsol}, ${\su}_{{{N}}-1}^F$ can be found while \eqref{vdef} yields ${\su}_{{{N}}}^F$. 
Finally, \eqref{ubulkK} provides the exact solution everywhere in 
half-plane {${{\mathfrak{S}}}_{\text{H}}$}. 
In particular,
by \eqref{vdef} and \eqref{LamK}, 
${\su}_{{{N}}}^F={\sv}_{{N}}^F+{\su}_{{{N}}-1}^F=(1+\frac{1}{{\mathtt{V}}_{{k}}}){\sv}_{{N}}^F$
and \eqref{ubulkK}{${}_1$}, \eqref{vNzsol}
yield
\begin{eqn}
{\su}_{{\mathtt{y}}}^F({{z}})={{\mathrm{A}}}{\mathtt{C}}_0\frac{{{z}} {{\mathtt{K}}}({{z}})}{{{z}}-{{z}}_{{P}}}(1+{{\mathtt{V}}_{{k}}({{z}})}^{-1}){{\lambda}}({{z}})^{{\mathtt{y}}-{{N}}}\quad\quad({\text{with }}{{\mathtt{y}}} \ge {{N}})
\label{uabovecrack}
\end{eqn}
{and 
additionally by \eqref{ubulkK}${}_2$ and
\eqref{u0uNn1K} (as well as using \eqref{fgK}${}_1$ and \eqref{fgK}${}_2$)
\begin{eqn}
{\su}_{{\mathtt{y}}}^F({{z}})&={{\mathrm{A}}}{\mathtt{C}}_0\frac{{{z}} {{\mathtt{K}}}({{z}})}{{{z}}-{{z}}_{{P}}}\big(\frac{(1+{{\upbeta}})f_{{{N}}-2}}{(\sQ+{{\upgamma}}-(1+{{\upbeta}})f_1)}({{{\lambda}}^{-2{{N}}+2}{{\lambda}}^{{\mathtt{y}}}-{{\lambda}}^{-{\mathtt{y}}}})-{{\lambda}}^{-{{N}}+1}({{{\lambda}}^{{\mathtt{y}}}-{{\lambda}}^{-{\mathtt{y}}}})\big)\\
&{{\mathtt{V}}_{{k}}({{z}})}^{-1}({{{\lambda}}^{-2{{N}}+2}-1})^{-1}\quad\quad\quad\quad({\text{with }}{{\mathtt{y}}}\in{\mathbb{Z}}_0^{{N}-1}).
\label{uwaveguidecrack}
\end{eqn}
}

\subsection{Rigid constraint}
\label{WHdiscreteslit}
For the lattice row at ${{\mathtt{y}}}={N}$, the equation satisfied by the scattered field is
\begin{eqn}
-{\upomega}^2{\su}_{{{\mathtt{x}}}, {{N}}}
={\triangle}{\su}_{{{\mathtt{x}}}, {\mathtt{y}}}, {{\mathtt{x}}}<0, {\mathtt{y}}={N}, 
\text{ and }
-{\upomega}^2{\su}_{{{\mathtt{x}}}, {{N}}}={\upomega}^2 {\su}^{\inc}_{{{\mathtt{x}}}, {{N}}}, {{\mathtt{x}}}\ge 0.
\label{discretesliteqn}
\end{eqn}
Clearly, $\sum\nolimits_{{\mathtt{x}}\in{\mathbb{Z}}}{{\mathcal{H}}}(-{\mathtt{x}}-1)\su_{{\mathtt{x}}+1,{N}}{z}^{-{\mathtt{x}}}
={z}\su_{{N}; -}+{z}\su_{0,{N}},$ and
$\sum\nolimits_{{\mathtt{x}}\in{\mathbb{Z}}}{{\mathcal{H}}}(-{\mathtt{x}}-1)\su_{{\mathtt{x}}-1,{N}}{z}^{-{\mathtt{x}}}
={z}^{-1}\su_{{N}; -}-\su_{-1,{N}}.$
Applying the 
Fourier transform 
{\eqref{discreteFT}}
to \eqref{discretesliteqn}, with ${\su}_{-1, {N}}$ as an unknown complex number, 
\beqan
{\sQ}{\su}_{{{N}}; -}
=-{{\mathtt{W}}}_{{N}}+{\su}_{{{N}}+1; -}+{\su}_{{{N}}-1; -}, 
\label{uNFeq}
\text{where }
{{\mathtt{W}}}_{{N}}={\su}_{-1, {N}}-{{z}} {\su}_{0, {N}}={\su}_{-1, {N}}+{{z}} {\su}^{\inc}_{0, {N}},\\
{\su}_{{{N}}; +}
=-\su^{\inc}_{{N}; +}, \quad\quad{\su}^{\inc}_{{N}; +}=\sum\limits_{{\mathtt{x}}=0}^\infty{{z}}^{-{\mathtt{x}}}{\su}^{\inc}_{{\mathtt{x}}, {{N}}}=\su^{\inc}_{0,{N}}\delta_{D+}({{z}} {z}_{{P}}^{-1}), \\
{\text{with }}
\su^{\inc}_{0,{N}}={{\mathrm{A}}}(e^{i{\upkappa}_y{N}}+c_Be^{-i{\upkappa}_y{N}}).
\label{uincN}
\eeqan
Analogous to the case of crack (with minor change in \eqref{ubulkK}${}_2$, replacing ${N}$ by ${N}+1$), extending the expression \eqref{u0uNn1K}, it is found that
\beqan
{\su}_{0}^F=\frac{(1+{{\upbeta}})f_{{{N}}-1}}{(\sQ+{{\upgamma}}-(1+{{\upbeta}})f_1)}{\su}_{{{N}}}^F, 
\quad\quad
\text{and }
\quad\quad
{\su}_{{{N}}-1}^F={\mathtt{V}}_{{c}}{\su}^F_{{{N}}},
\quad
{\su}_{{{N}}+1}^F={{\lambda}}\su^F_{{N}},
\label{u1NC}\text{ where }\\
{\mathtt{V}}_{{c}}=\frac{(1+{{\upbeta}})f_{{{N}}-1}}{(\sQ+{{\upgamma}}-(1+{{\upbeta}})f_1)}f_{{{N}}-1}+f_1,
\label{LamC}
f_1=\frac{{{\lambda}}^{-{{N}}+1}-{{\lambda}}^{{{N}}-1}}{{{\lambda}}^{-2{{N}}}-1}{{\lambda}}^{-{{N}}},
f_{{{N}}-1}=\frac{{{\lambda}}^{-1}-{{\lambda}}}{{{\lambda}}^{-2{{N}}}-1}{{\lambda}}^{-{{N}}}.
\eeqan
Using 
\eqref{uNFeq}
and \eqref{u1NC}, a {{WH}} equation is found for 
\begin{eqn}
\mathscrpring_{{{N}}; \pm}={\su}_{{{N}}-1; \pm}+{\su}_{{{N}}+1; \pm},
\label{sumuNNp}
\end{eqn}
as
\beqan
{\mathtt{L}}\mathscrpring_{{{N}}; +}+\mathscrpring_{{{N}}; -}
=(1-{\mathtt{L}})({{\mathtt{W}}}_{{N}}-{\sQ}{\su}_{{{N}}; +}),
\label{WHC}
\text{ where }
{\mathtt{L}}
=\frac{{\sQ}}{{\lambda}^{-1}-{\mathtt{V}}_{{c}}}
={\mathfrak{F}}_{{c}}{{\mathtt{L}}}_{{c}}^{-1},\\
\text{with the structure factor }
{\mathfrak{F}}_{{c}}({z}; {{\upbeta}}, {{\upgamma}}, {{N}})=1+{\mathtt{C}}_B({\lambda}){{\lambda}}^{2 {N}},
\label{defFslit}
\eeqan
employing definition of ${\mathtt{C}}_B$ \eqref{defCB} and ${{\mathtt{L}}}_{{c}}$ ($=\sr\sh/\sQ$) \cite{Bls1}. 
{The {{WH}} equation \eqref{WHC} is also posed on an annulus
${\mathscr{A}}$ in the complex plane same as that \eqref{annAAL} employed earlier for the crack.}
As ${N}\to\infty$, the strip lemma of \cite{Bls9s} holds in a manner similar to that stated before for the case of crack, {see Remark \ref{limitN}}.

Using the multiplicative factorization ${{\mathtt{L}}}_{{}}={{\mathtt{L}}}_{{}+}{{\mathtt{L}}}_{{}-}$, the {{WH}} equation \eqref{WHC} becomes
\begin{eqn}
{\mathtt{L}}_+\mathscrpring_{{{N}}; +}+{\mathtt{L}}_-^{-1}\mathscrpring_{{{N}}; -}
={\mathtt{C}}, 
\label{WHCn}
\text{with }
{\mathtt{C}}=({\mathtt{L}}_-^{-1}-{\mathtt{L}}_+)({{\mathtt{W}}}_{{N}}+{\sQ}{\su}^{\inc}_{{{N}}; +}).
\end{eqn}
An additive factorization \cite{Noble} of right hand side, i.e., ${{\mathtt{C}}}={{\mathtt{C}}}_{+}+{{\mathtt{C}}}_{-},$ 
on ${\mathscr{A}}$, holds with
 \begin{eqn}
{{\mathtt{C}}}_\pm({{z}})&=\mp{\su}_{-1, {N}}({{\mathtt{L}}}_{\pm}^{\pm1}({{z}})-{\overline{l}}_{-0})\mp{{z}} {\su}^{\inc}_{0, {N}}({{\mathtt{L}}}_{\pm}^{\pm1}({{z}})-{l_{}}_{+0})\mp\su^{\inc}_{0,{N}}\delta_{D+}({{z}} {{z}}_{{P}}^{-1})\\&\big({\sQ}({{z}}){{\mathtt{L}}}_{\pm}^{\pm1}({{z}})-{\sQ}({{z}}_{{P}}){{\mathtt{L}}}_-^{-1}({{z}}_{{P}})+{\overline{l}}_{-0}({{z}}^{-1}-{{z}}_{{P}}^{-1})+{{l}}_{+0}({{z}}-{{z}}_{{P}})\big), 
\label{CpmC}
\end{eqn}
with 
\begin{eqn}
{l_{}}_{+0}=\lim_{{z}\to\infty}{{\mathtt{L}}}_+({{z}})\text{ and }{\overline{l}}_{-0}=\lim_{{z}\to0}{{{\mathtt{L}}}^{-1}_{}}_-({{z}}).
\end{eqn}
The function ${{\mathtt{C}}}_+({{z}})$ (resp. ${{\mathtt{C}}}_-({{z}})$) is analytic at ${{z}}\in{\mathbb{C}}$ such that $|{{z}}|>\max\{{{\mathrm{R}}}_+, {{\mathrm{R}}}_{L}\}$ (resp. $|{{z}}|<\min\{{{\mathrm{R}}}_-, {{\mathrm{R}}}_{L}^{-1}\}$). 
An application of the Liouville's theorem (using elementary estimates on the kernel as well as the boundedness of {the sequence corresponding to} $\mathscrpring_{{N}}$ \eqref{sumuNNp}) leads to the solution of \eqref{WHC},
\begin{eqn}
{\mathscrpring}_{{N}; \pm}({{z}})&= {{\mathtt{C}}}_\pm({{z}}){{\mathtt{L}}}_{\pm}({{z}})^{\mp1}, 
\quad{{z}}\in{\mathbb{C}}, |{{z}}|>\bfrac{\max}{\min}\{{{\mathrm{R}}}_\pm, {{\mathrm{R}}}_{L}^{\pm1}\}.
\label{wNpmC}
\end{eqn}
Using \eqref{uNFeq} and \eqref{uNFeq_altinc}, the expression for ${\su}_{{N}; +}$ can be found 
by incorporating minor changes in the expressions and manipulations detailed for the infinite lattice in \cite{Bls1}. 
Indeed, 
as detailed in the 
supplementary 1, it is found that
\begin{eqn}
{\su}^{\totwave}_{-1, {N}}=-\su^{\inc}_{0,{N}}\frac{{{z}}_{\sq}}{{{z}}_{\sq}-{{z}}_{{P}}}\frac{{\sQ}({{z}}_{{P}})}{{\overline{l}}_{-0}{{\mathtt{L}}}_-({{z}}_{{P}})}.
\label{un1Ntot}
\end{eqn}
By \eqref{uNFeq},
\begin{eqn}
{\su}^F_{{N}}
&=\frac{{{\mathtt{L}}}_{-}({{z}})}{\sQ({z})}(-\su^{\inc}_{0,{N}}\delta_{D+}({{z}} {{z}}_{{P}}^{-1}){\sQ}({{z}}_{{P}}){{\mathtt{L}}}_-^{-1}({{z}}_{{P}})-\su^{\inc}_{0,{N}}{\overline{l}}_{-0}{{z}}_{{P}}^{-1}+{\su}_{-1, {N}}(-{\overline{l}}_{-0})).
\end{eqn}
Using 
$\sQ({z})={{z}}_{\sq}^{-1}(1-{{z}}_{\sq}{{z}})(1-{{z}}_{\sq}{{z}}^{-1})$ \cite{Bls1},
and \eqref{un1Ntot},
\begin{eqn}
{\su}_{{N}}^F({{z}})={{\mathrm{A}}}{\mathtt{C}}_0\frac{{{z}}{{\mathtt{K}}}({{z}})}{{{z}}-{{z}}_{{P}}}, 
\text{where }
{{\mathtt{K}}}({{z}}){:=}\frac{{{\mathtt{L}}}_{-}({{z}})}{(1-{{z}}_{\sq}{{z}})}, {{z}}\in{{\mathscr{A}}},\\
{\mathtt{C}}_0{:=}{{\mathrm{A}}}^{-1}\su^{\inc}_{0,{N}}{{z}}_{{P}}{{\sQ}({{z}}_{{P}}){{\mathtt{L}}}^{-1}_-({{z}}_{{P}})}{{z}}_{\sq}({{z}}_{\sq}-{{z}}_{{P}})^{-1}
\in{\mathbb{C}}.
\label{uNzsol}
\end{eqn}
Observe that ${{\mathrm{A}}}{\mathtt{C}}_0$ happens to be same as $-{\su}^{\totwave}_{-1, {N}}{\overline{l}}_{-0}$ by a recall of \eqref{un1Ntot}.

In the case of {\em incidence from the waveguide}, the scattering occurs due to the unconstrained sites ahead of the upper boundary of waveguide.
In contrast to \eqref{uinc}, the incident wave is given by \eqref{uinc_altinc}, where ${{a}}_{({{{\kappa}}^{\inc}}){{\mathtt{y}}}}$ refers to the eigenmode representing a propagating wave in the lattice waveguide formed by 
half-plane boundary and the rigid constraint. Notice that ${{a}}_{({{{\kappa}}^{\inc}}){{\mathtt{y}}}}$ automatically satisfies the fixed boundary condition at ${\mathtt{y}}={N}.$
\eqref{uNFeq} is replaced by
\begin{eqn}
{\sQ}{\su}_{{{N}}; -}=-{{\mathtt{W}}}_{{N}}+{\su}_{{{N}}+1; -}+{\su}_{{{N}}-1; -}+{\su}^{\inc}_{{{N}}-1; -}, 
\label{uNFeq_altinc}
\end{eqn}
while $-{\upomega}^2{\su}_{{{\mathtt{x}}}, {{N}}}=0, {{\mathtt{x}}}\ge 0.$ Let ${\mathscrpring}^{\inc}_{{N}; -}={\su}^{\inc}_{{{N}}-1; -}+{\su}^{\inc}_{{{N}}+1; -}={\su}^{\inc}_{{{N}}-1; -}+0={\su}^{\inc}_{{{N}}-1; -}.$ Note that ${{\mathtt{W}}}_{{N}}={\su}_{-1, {N}}+{{z}} {\su}^{\inc}_{0, {N}}={\su}_{-1, {N}}$.
With ${\mathtt{C}}=({{\mathtt{W}}}_{{N}}-{\mathscrpring}^{\inc}_{{N}; -})({{\mathtt{L}}}_{-}^{-1}-{{\mathtt{L}}}_{+})$ in place of ${\mathtt{C}}$, 
the (same) 
equation \eqref{WHCn} results;
its additive factorization holds with 
\begin{eqn}
{{\mathtt{C}}}_{\pm}({{z}})&=\mp{\su}_{-1, {N}}({{\mathtt{L}}}_\pm^{\pm1}({{z}})-{\overline{l}}_{-0})\pm{\mathscrpring}^{\inc}_{0,{N}}\delta_{D-}({{z}} {{z}}_{{P}}^{-1})
\big({{\mathtt{L}}}_{\pm}^{\pm1}({{z}})-{{\mathtt{L}}}_+({{z}}_{{P}})\big),
\label{CpmC_altinc}
\end{eqn}
where ${\overline{l}}_{-0}=\lim_{{{z}}\to0}{{\mathtt{L}}}_-^{-1}({{z}})$. 
Finally, the solution of \eqref{WHCn} is written as \eqref{wNpmC}.
Also, as detailed in the 
supplementary 1, it is found that
\begin{eqn}
{\su}^{\totwave}_{-1, {N}}=-\su^{\inc}_{0,{N}-1}\frac{{{z}}_{\sq}}{{{z}}_{\sq}-{{z}}_{{P}}}\frac{{{\mathtt{L}}}_+({{z}}_{{P}})}{{\overline{l}}_{-0}}.
\label{un1Ntot_altinc}
\end{eqn}
In the case of {\em incidence from the waveguide}, \eqref{uNzsol} follows with
\begin{eqn}
{\mathtt{C}}_0{:=}{{\mathrm{A}}}^{-1}\su^{\inc}_{0,{N}-1}{{z}}_{{P}}{{\mathtt{L}}}_+({{z}}_{{P}}){{z}}_{\sq}({{z}}_{\sq}-{{z}}_{{P}})^{-1}\in{\mathbb{C}}.
\end{eqn}

By using \eqref{LamC} and \eqref{vNzsol}, as well as \eqref{u1NC}, ${\su}_{{{N}}-1}^F$ and ${\su}_{{{N}}+1}^F$ can be found.
In fact, an analogue of \eqref{ubulkK} provides the exact solution everywhere. 
In particular,
\eqref{ubulkK}${}_1$, \eqref{uNzsol}
yields
\begin{eqn}
{\su}_{{\mathtt{y}}}^F({{z}})={{\mathrm{A}}}{\mathtt{C}}_0({{{z}} {{\mathtt{K}}}({{z}})}/({{{z}}-{{z}}_{{P}}})){{\lambda}}({{z}})^{{\mathtt{y}}-{{N}}}\quad\quad({\text{with }}{{\mathtt{y}}} \ge {{N}})
\label{uaboveconstraint}
\end{eqn}
{and 
additionally by \eqref{ubulkK}${}_2$ and
\eqref{u1NC}${}_1$ (as well as using \eqref{LamC}${}_2$ and \eqref{LamC}${}_3$)
\begin{eqn}
{\su}_{{\mathtt{y}}}^F({{z}})&={{\mathrm{A}}}{\mathtt{C}}_0\frac{{{z}} {{\mathtt{K}}}({{z}})}{{{z}}-{{z}}_{{P}}}\big(\frac{(1+{{\upbeta}})f_{{{N}}-1}}{(\sQ+{{\upgamma}}-(1+{{\upbeta}})f_1)}({{{\lambda}}^{-2{{N}}}{{\lambda}}^{{\mathtt{y}}}-{{\lambda}}^{-{\mathtt{y}}}})-{{\mathtt{V}}_{{c}}}{{\lambda}}^{-{{N}}}({{{\lambda}}^{{\mathtt{y}}}-{{\lambda}}^{-{\mathtt{y}}}})\big)\\
&({{{\lambda}}^{-2{{N}}}-1})^{-1}\quad\quad\quad\quad\quad\quad({\text{with }}{\mathtt{y}}\in{\mathbb{Z}}_0^{{N}-1}).
\label{uwaveguideconstraint}
\end{eqn}
}
{Above is not surprising,} since the expression of ${\su}_{{{N}}-1; \pm}+{\su}_{{{N}}+1; \pm}$ can be used 
to determine ${\su}_{{{N}}; -}$ by \eqref{uNFeq}, {so that} the problem is solved completely by \eqref{u1NC}.

\section{Far field approximation in the reduced half-plane problem}
\label{farfield}
\subsection{Far field approximation in the bulk lattice}
In either case, i.e. crack or rigid constraint, ${\su}_{{\mathtt{x}}, {\mathtt{y}}}$ is eventually determined by inverse 
Fourier transform, 
\begin{eqn}
{\su}_{{\mathtt{x}}, {\mathtt{y}}}=&\frac{1}{2\pi i}\oint_{{{\mathcal{C}}}_{{z}}}{\su}_{{\mathtt{y}}}^{F}({{z}}){{z}}^{{\mathtt{x}}-1}d{{z}}, 
\quad\quad
{({\mathtt{x}}, {\mathtt{y}})\in{{\mathbb{Z}}}^2_{\text{H}}},
\label{umnsol_sq}
\end{eqn}
where ${{\mathcal{C}}}_{{z}}$ is a rectifiable, closed, counterclockwise contour (an appropriately dented {contour, most of which coincides with} the unit circle $\mathbb{T}\subset{\mathbb{C}}$ in case the limit ${\upomega}_2\to0^+$ is considered) in the annulus ${{\mathscr{A}}}$ (recall \eqref{annAAL} {and Remark \ref{annuluswginc}}). 
Following the analysis of \cite{Bls0,Bls1}, with 
${{z}}=e^{-i{\upxi}}$,
{
\begin{eqn}
{\mathtt{x}}={R}\cos{\theta},
{\mathtt{y}}=-{\tfrac{1}{2}}(1-{{\,\gimel\,}})
+{R}\sin{\theta},
\label{polar_sq}\end{eqn}
}
for the incidence from the bulk lattice, the expression \eqref{umnsol_sq} can be rewritten, 
in case of rigid constraint, using \eqref{uNzsol} and \eqref{ubulkK}${}_1$ for ${\mathtt{y}}\ge{N}$, as
\begin{subequations}
\begin{eqn}
{\su}_{{{\mathtt{x}}}, {{\mathtt{y}}}}=-\frac{1}{2\pi}{{\mathrm{A}}}{\mathtt{C}}_0\int_{{{\mathcal{C}}}_{{{\upxi}}}}\frac{{{\mathtt{K}}}(e^{-i{{\upxi}}})e^{i{R}\upphi({{\upxi}})}}{e^{i({\upxi}-{\upxi}_{{P}})}-1}e^{-i({N}+{\tfrac{1}{2}}(1-{{\,\gimel\,}})){\upeta}({{\upxi}})}d{{\upxi}},
\label{umnsolrtC}
\end{eqn}
while, for the crack, 
using \eqref{vNzsol} and \eqref{ubulkK}${}_1$, 
for ${\mathtt{y}}\ge{N}$,
\begin{eqn}
{\su}_{{{\mathtt{x}}}, {{\mathtt{y}}}}=-\frac{1}{2\pi}{{\mathrm{A}}}{\mathtt{C}}_0\int_{{{\mathcal{C}}}_{{{\upxi}}}}(1+\frac{1}{{\mathtt{V}}(e^{-i{{\upxi}}})})\frac{{{\mathtt{K}}}(e^{-i{{\upxi}}})e^{i{R}\upphi({{\upxi}})}}{e^{i({\upxi}-{\upxi}_{{P}})}-1}e^{-i({N}+{\tfrac{1}{2}}(1-{{\,\gimel\,}})){\upeta}({{\upxi}})}d{{\upxi}}.
\label{umnsolrtK}
\end{eqn}
\label{umnsolrt}
\end{subequations}
In \eqref{umnsolrt}, ${{\mathcal{C}}}_{{{\upxi}}}$ is a contour (oriented along increasing ${\upxi}_1$) which
lies in the strip ${{\mathscr{S}}}=\{{\upxi}\in{\mathbb{C}}: 
{\upxi}_1\in[-\pi+\pi {\mathcal{H}}({\upomega}-2){\mathcal{H}}(2\sqrt{2}-{\upomega}), \pi+\pi {\mathcal{H}}({\upomega}-2){\mathcal{H}}(2\sqrt{2}-{\upomega})], 
-{\upkappa}_2<{\upxi}_2<{\upkappa}_2\cos{\Theta}\},$ ${\upxi}_{{P}}=-{\upkappa}_x$, and
$\upphi({\upxi})={\upeta}({\upxi})\sin{\theta}-{\upxi}\cos{\theta}, {\upeta}({\upxi})=-i\log{{\lambda}}(e^{-i{\upxi}}), 
{\upxi}\in {{\mathscr{S}}}.$
Eventually, by an application of the results provided by \cite{Bls0,Bls1},
(with ${\upxi}={\upxi}_{{S}}$ as the saddle point 
of $\upphi$ on ${\mathcal{C}}_{{\upxi}}$) the far-field approximation, 
for the case of rigid constraint, is
${\su}_{{\mathtt{x}}, {\mathtt{y}}}\sim{\su}_{{\mathtt{x}}, {\mathtt{y}}}|_{{S}}+{\su}_{{\mathtt{x}}, {\mathtt{y}}}|_{{P}}$
\begin{subequations}
where
\beqan
{\su}_{{\mathtt{x}}, {\mathtt{y}}}|_{{S}}&\sim&-{{\mathrm{A}}}{\mathtt{C}}_{0}{{\mathtt{K}}}({{z}}_{{S}})\frac{1+i{\text{\rm sgn}}({\upeta}''({\upxi}_{{S}}))}{2\sqrt{\pi}}\frac{e^{i{R}({\upeta}({\upxi}_{{S}})\sin{\theta}-{\upxi}_{{S}}\cos{\theta})}}{({R}|{\upeta}''({\upxi}_{{S}})|\sin{\theta})^{{\tfrac{1}{2}}}}\frac{e^{-i({N}+{\tfrac{1}{2}}(1-{{\,\gimel\,}})){\upeta}({\upxi}_{{S}})}}{{{z}}_{{P}}{{z}}_{{S}}^{-1}-1},
\label{statphase}\\
{\su}_{{\mathtt{x}}, {\mathtt{y}}}|_{{P}}&=&{\su}_{{\mathtt{x}}, {\mathtt{y}}}^{\refwave}{{{\mathcal{H}}}({\theta}_{\refwave}-{\theta})},
\label{umnpole}
\text{and }\\
{\su}_{{\mathtt{x}}, {\mathtt{y}}}^{\refwave}&=&{{\mathrm{A}}}{\mathtt{C}}_0{{z}}_{{P}}{{\mathtt{K}}}({{z}}_{{P}}){\lambda}({{z}}_{{P}})^{{\mathtt{y}}-{N}}{{z}}_{{P}}^{{\mathtt{x}}-1}
\label{urefC}
\eeqan
\end{subequations}
while for the case of crack, there is a pre-factor $(1+{{\mathtt{V}}({{z}}_{{S}})}^{-1})$ in \eqref{statphase} and
\begin{eqn}
{\su}_{{\mathtt{x}}, {\mathtt{y}}}^{\refwave}&=(1+{{\mathtt{V}}({{z}}_{{P}})}^{-1}){{\mathrm{A}}}{\mathtt{C}}_0{{z}}_{{P}}{{\mathtt{K}}}({{z}}_{{P}}){\lambda}({{z}}_{{P}})^{{\mathtt{y}}-{N}}{{z}}_{{P}}^{{\mathtt{x}}-1}.
\label{urefK}
\end{eqn}
Equations \eqref{urefC} and \eqref{urefK} can be simplified further to obtain
${\su}_{{\mathtt{x}}, {\mathtt{y}}}^{\refwave}=-\su^{\inc}_{0,{N}}
e^{i{\upkappa}_x {\mathtt{x}}+i{\upkappa}_y({\mathtt{y}}-{N})}$
and
${\su}_{{\mathtt{x}}, {\mathtt{y}}}^{\refwave}
=-{\sv}^{\inc}_{0,{N}}(1-e^{-i{\upkappa}_y})^{-1}e^{i{\upkappa}_x {\mathtt{x}}+i{\upkappa}_y({\mathtt{y}}-{N})},$
where
${\sv}^{\inc}_{0,{N}}$ is given by \eqref{svinc0N}.

Similar expressions 
can be obtained for {\em incidence from the waveguide}; the details are omitted.

\subsection{Far field approximation in the lattice waveguide}
Due to the vanishing of the diffracted wave field in the immediate vicinity farther behind the crack or rigid constraint tip, it is natural to seek an expansion of the expression of $\sv_{{\mathtt{x}}, {N}}$ in case of crack and $\mathscrpring_{{\mathtt{x}}, {N}}$ in case of rigid constraint, as ${\mathtt{x}}\to\infty.$
Noting the absence of the contribution of $\su_{{\mathtt{x}}, {N}}$ and $\su_{{\mathtt{x}}, {N}+1}$ in the respective cases, the function $\sv_{{N};+}$ and $\mathscrpring_{{N};+}$ play the pivotal role. For the case of crack, using plus ($+$) part of \eqref{vNpmK}, with its counterpart in \eqref{CpmK} for incidence from the bulk lattice ({denoted by }${{\mathfrak{s}}}={\tt{B}}$) while that in \eqref{CpmK_altinc} for {\em incidence from the waveguide} ({denoted by }${{\mathfrak{s}}}={\tt{W}}$), i.e.,
it is found that
\begin{eqn}
{\sv}_{{{N}}; +}
&={\sv}^{\inc}_{0,{N}}({{\mathtt{L}}}_{+}^{-1}({{z}}){{\mathtt{L}}}^{-1}_{{}-}({{z}}_{{P}})-1)\delta_{D+}({{z}} {{z}}_{{P}}^{-1})\delta_{{{\mathfrak{s}}},{\tt{B}}}\\&
-{\sv}^{\inc}_{0,{N}}({{\mathtt{L}}}_{+}^{-1}({{z}}){{{\mathtt{L}}}_{+}({{z}}_{{P}})}-1)\delta_{D-}({{z}} {{z}}_{{P}}^{-1})\delta_{{{\mathfrak{s}}},{\tt{W}}}.
\label{vNplus}
\end{eqn}
Using the inverse 
Fourier transform 
{\eqref{inversediscreteFT}}
and residue calculus \cite{Ablowitz}, noting that ${\sv}_{{\mathtt{x}}, {{N}}}\sim -{\su}_{{\mathtt{x}}, {{N}}-1}$ as ${\mathtt{x}}\to\infty$, \eqref{vNplus} yields 
\begin{eqn}
{\su}_{{\mathtt{x}},{N}-1}&\sim -{\sv}^{\inc}_{0,{N}}(({\mathtt{L}}^{-1}({{z}}_{{P}})-1){{z}}_{{P}}^{{\mathtt{x}}}+\sum\nolimits_{{{\mathtt{L}}}_{+}({{z}})=0}\frac{1}{{z}-{z}_{{P}}}\frac{{{\mathtt{L}}}^{-1}_{{}-}({{z}}_{{P}})}{{{\mathtt{L}}}'_{+}({{z}})}{z}^{{\mathtt{x}}})\delta_{{{\mathfrak{s}}},{\tt{B}}}\\
&-{\sv}^{\inc}_{0,{N}}\sum\nolimits_{{{\mathtt{L}}}_{+}({{z}})=0}\frac{1}{{z}-{z}_{{P}}}\frac{{\mathtt{L}}_+({{z}}_{{P}})}{{{\mathtt{L}}}'_{+}({{z}})}{z}^{{\mathtt{x}}}\delta_{{{\mathfrak{s}}},{\tt{W}}}.
\label{uNn1K_far}
\end{eqn}
For the case of rigid constraint, using plus ($+$) part of \eqref{wNpmC}, with its counterpart in \eqref{CpmC} for incidence from the bulk lattice (${{\mathfrak{s}}}={\tt{B}}$) while that in \eqref{CpmC_altinc} for {\em incidence from the waveguide} (${{\mathfrak{s}}}={\tt{W}}$), i.e.,
it is found that
\begin{eqn}
{\mathscrpring}_{{N}; +}({{z}})
&={\su}^{\inc}_{0, {N}}(\frac{{{z}}_{\sq}{\sQ}({{z}}_{{P}})}{{{z}}_{\sq}-{{z}}_{{P}}}\frac{{{\mathtt{L}}}_-^{-1}({{z}}_{{P}})}{{\overline{l}}_{-0}}-\frac{{{z}}{\sQ}({{z}})}{{{z}}-{{z}}_{{P}}}+\frac{-{{z}}_{{P}}({z}-{{z}}_{\sq}){\sQ}({{z}}_{{P}})}{({{z}}-{{z}}_{{P}})({{z}}_{\sq}-{{z}}_{{P}})}\frac{{{\mathtt{L}}}_-^{-1}({{z}}_{{P}})}{{{\mathtt{L}}}_{+}({{z}})})\delta_{{{\mathfrak{s}}},{\tt{B}}}\\
&+\su^{\inc}_{0,{N}-1}(\frac{{{z}}_{\sq}}{{{z}}_{\sq}-{{z}}_{{P}}}\frac{{{\mathtt{L}}}_+({{z}}_{{P}})}{{\overline{l}}_{-0}}-\frac{{z}}{{z}-{{z}}_{{P}}}+\frac{-{{z}}_{{P}}({z}-{{z}}_{\sq})}{({{z}}-{{z}}_{{P}})({{z}}_{\sq}-{{z}}_{{P}})}\frac{{{\mathtt{L}}}_+({{z}}_{{P}})}{{{\mathtt{L}}}_{+}({{z}})})\delta_{{{\mathfrak{s}}},{\tt{W}}}.
\label{uNn1plus}
\end{eqn}
Using the inverse 
Fourier transform 
{\eqref{discreteFT}}
and residue calculus, noting that ${\mathscrpring}_{{\mathtt{x}}, {{N}}}\sim {\su}_{{\mathtt{x}}, {{N}}-1}$ as ${\mathtt{x}}\to\infty$, \eqref{vNplus} and $\sQ_\pm({z})={{z}}_{\sq}^{-1/2}(1-{{z}}_{\sq}{{z}}^{\mp1})$
yields 
\begin{eqn}
{\su}_{{\mathtt{x}},{N}-1}
&\sim {\su}^{\inc}_{0,{N}}({\sQ}({{z}}_{{P}})(\frac{1}{{\mathtt{L}}({{z}}_{{P}})}-1){{z}}_{{P}}^{{\mathtt{x}}}+\sum\nolimits_{{{\mathtt{L}}}_{+}({{z}})=0}\frac{1}{{z}-{z}_{{P}}}\frac{{\sQ}_-({{z}}_{{P}})}{{{\mathtt{L}}}_-({{z}}_{{P}})}\frac{\sQ_+({z})}{{{\mathtt{L}}}'_{+}({{z}})}{z}^{{\mathtt{x}}})\delta_{{{\mathfrak{s}}},{\tt{B}}}\\
&+{\su}^{\inc}_{0,{N}-1}\sum\nolimits_{{{\mathtt{L}}}_{+}({{z}})=0}\frac{1}{{z}-{z}_{{P}}}\frac{{{\mathtt{L}}}_+({{z}}_{{P}})}{\sQ_+({z}_{{P}})}\frac{\sQ_+({z})}{{{\mathtt{L}}}'_{+}({{z}})}{z}^{{\mathtt{x}}}\delta_{{{\mathfrak{s}}},{\tt{W}}}.
\label{uNn1C_far}
\end{eqn}
Using
the expression of total wave field corresponding to \eqref{uNn1K_far} and \eqref{uNn1C_far}, the unknown coefficients in its eigenmode expansion, 
deep inside the waveguide, can be obtained in a straightforward manner based on orthogonality of modes \cite{Bls9} (denoted by ${{a}_{({\kappa})\cdot}}$). Finally, a far-field expansion of total wave field (with ${\mathtt{y}}\in{\mathbb{Z}}_0^{{N}-1}$) is found to be, for the case of crack,
\begin{eqn}
\su^{\totwave}_{{\mathtt{x}}, {\mathtt{y}}}&\sim -{\sv}^{\inc}_{0,{N}}({{\mathtt{L}}}^{-1}_{{}-}({{z}}_{{P}})\delta_{{{\mathfrak{s}}},{\tt{B}}}+{\mathtt{L}}_+({{z}}_{{P}})\delta_{{{\mathfrak{s}}},{\tt{W}}})\hspace{-.1in}\sum_{{{\mathtt{L}}}_{+}({{z}})=0}\frac{{a}_{({\kappa}){\mathtt{y}}}}{{a}_{({\kappa}){N}-1}}\frac{1}{{z}-{z}_{{P}}}\frac{{z}^{{\mathtt{x}}}}{{{\mathtt{L}}}'_{+}({{z}})},
\label{uxywaveguideK_far}
\end{eqn}
and, for the case of rigid constraint,
\begin{eqn}
\su^{\totwave}_{{\mathtt{x}}, {\mathtt{y}}}&\sim({\su}^{\inc}_{0,{N}}\frac{{\sQ}_-({{z}}_{{P}})}{{{\mathtt{L}}}_-({{z}}_{{P}})}\delta_{{{\mathfrak{s}}},{\tt{B}}}+{\su}^{\inc}_{0,{N}-1}\frac{{{\mathtt{L}}}_+({{z}}_{{P}})}{\sQ_+({z}_{{P}})}\delta_{{{\mathfrak{s}}},{\tt{W}}})\hspace{-.1in}\sum_{{{\mathtt{L}}}_{+}({{z}})=0}\frac{{a}_{({\kappa}){\mathtt{y}}}}{{a}_{({\kappa}){N}-1}}\frac{1}{{z}-{z}_{{P}}}\frac{\sQ_+({z}){z}^{{\mathtt{x}}}}{{{\mathtt{L}}}'_{+}({{z}})}.
\label{uxywaveguideC_far}
\end{eqn}
Indeed, \eqref{uxywaveguideK_far} and \eqref{uxywaveguideC_far} can also be obtained directly by using the expression \eqref{ubulkK}${}_2$.

\section{Back to the problem involving a pair of parallel defects}
\label{paircrackslit}
Reverting back to the main motivation for this paper, i.e., the analysis of diffraction of wave incident from the bulk lattice \eqref{uincB} by a pair of parallel cracks or rigid constraints, the wave field (diffracted) modulo the reflected wave from the geometrically reduced problem can be superposed 
in order to construct an exact solution.

For the purpose of symbolic convenience, suppose that the scattered wave field 
for four choices of ${{\upbeta}}, {{\upgamma}}$, i.e., cases H\ref{cond1}--H\ref{cond4}, 
are denoted by
\begin{eqn}
\su^{\scawave}_{{\mathtt{x}}, {\mathtt{y}}}({{\mathrm{A}}}, {\upkappa}_x, {\upkappa}_y; {{\upbeta}}, {{\upgamma}}; {N}, {k}), 
\quad
\su^{\scawave}_{{\mathtt{x}}, {\mathtt{y}}}({{\mathrm{A}}}, {\upkappa}_x, {\upkappa}_y; {{\upbeta}}, {{\upgamma}}; {N}, {c}),
\label{udiffracsymbol}
\end{eqn}
where the former corresponds to a crack located at ${\mathtt{y}}={N}, {N}-1$ and the latter corresponds to a rigid constraint located at ${\mathtt{y}}={N},$ while the boundary of half-plane (of type depending on ${{\upbeta}}, {{\upgamma}}$) is located at ${\mathtt{y}}=0$ in both cases and the expression of 
incident wave remains the same (equal to constant ${{\mathrm{A}}}$ at $(0, 0)$, 
without the reflected wave contribution).

{At this point, recall \S\ref{georeduction}; in particular, Equations \eqref{uincBevenall} and \eqref{uincBoddall} which decompose the incident wave into even-symmetric and odd-symmetric components.}

\subsection{Even separation: $2{N}$}
In this case only the cases H\ref{cond2} and H\ref{cond3} are possible.
The parity bit is ${{\,\gimel\,}}=0$.
It is easy to see that the scattered wave field solution is given by
$\su^{\scawave}_{{\mathtt{x}}, {\mathtt{y}}}=\su^{\scawave}_{{\mathtt{x}}, {\mathtt{y}}}({\tfrac{1}{2}}{{\mathrm{A}}}, {\upkappa}_x, {\upkappa}_y; \text{for case } H\ref{cond2})+\su^{\scawave}_{{\mathtt{x}}, {\mathtt{y}}}({\tfrac{1}{2}}{{\mathrm{A}}}, {\upkappa}_x, {\upkappa}_y; \text{for case }H\ref{cond3}).$
In particular, for the crack problem,
\begin{eqn}
{\su^{\scawave}_{{\mathtt{x}}, {\mathtt{y}}}=\su^{\scawave}_{{\mathtt{x}}, {\mathtt{y}}}({\tfrac{1}{2}}{{\mathrm{A}}}, {\upkappa}_x, {\upkappa}_y; 0, -1; {N}, {k})+\su^{\scawave}_{{\mathtt{x}}, {\mathtt{y}}}({\tfrac{1}{2}}{{\mathrm{A}}}, {\upkappa}_x, {\upkappa}_y; 0, 1; {N}, {k}),}
\label{crackevensep}
\end{eqn}
and for the rigid constraint problem,
\begin{eqn}
{\su^{\scawave}_{{\mathtt{x}}, {\mathtt{y}}}=\su^{\scawave}_{{\mathtt{x}}, {\mathtt{y}}}({\tfrac{1}{2}}{{\mathrm{A}}}, {\upkappa}_x, {\upkappa}_y; 0, -1; {N}, {c})+\su^{\scawave}_{{\mathtt{x}}, {\mathtt{y}}}({\tfrac{1}{2}}{{\mathrm{A}}}, {\upkappa}_x, {\upkappa}_y; 0, 1; {N}, {c}).}
\label{slitevensep}
\end{eqn}

\subsection{Odd separation: $2{N}-1$}
In this case only the cases H\ref{cond1} and H\ref{cond4} are possible. 
The parity bit is ${{\,\gimel\,}}=1$.
Due to the choice of boundary location in case of H\ref{cond1}, the value of ${N}$ needs to mapped properly.
It is easy to see that the scattered wave field solution is given by
$\su^{\scawave}_{{\mathtt{x}}, {\mathtt{y}}}=\su^{\scawave}_{{\mathtt{x}}, {\mathtt{y}}-1}({\tfrac{1}{2}}{{\mathrm{A}}}e^{i{\upkappa}_y}, {\upkappa}_x, {\upkappa}_y; \text{for case }H\ref{cond1})+\su^{\scawave}_{{\mathtt{x}}, {\mathtt{y}}}({\tfrac{1}{2}}{{\mathrm{A}}}, {\upkappa}_x, {\upkappa}_y; \text{for case }H\ref{cond4}).$
In particular, for the crack problem,
\begin{eqn}
{\su^{\scawave}_{{\mathtt{x}}, {\mathtt{y}}}=\su^{\scawave}_{{\mathtt{x}}, {\mathtt{y}}-1}({\tfrac{1}{2}}{{\mathrm{A}}}e^{i{\upkappa}_y}, {\upkappa}_x, {\upkappa}_y; 0, 0; {N}-1, {k})+\su^{\scawave}_{{\mathtt{x}}, {\mathtt{y}}}({\tfrac{1}{2}}{{\mathrm{A}}}, {\upkappa}_x, {\upkappa}_y; 1, 0; {N}, {k}),}
\label{crackoddsep}\end{eqn}
and for the rigid constraint problem,
\begin{eqn}
{\su^{\scawave}_{{\mathtt{x}}, {\mathtt{y}}}=\su^{\scawave}_{{\mathtt{x}}, {\mathtt{y}}-1}({\tfrac{1}{2}}{{\mathrm{A}}}e^{i{\upkappa}_y}, {\upkappa}_x, {\upkappa}_y; 0, 0; {N}-1, {c})+\su^{\scawave}_{{\mathtt{x}}, {\mathtt{y}}}({\tfrac{1}{2}}{{\mathrm{A}}}, {\upkappa}_x, {\upkappa}_y; 1, 0; {N}, {c}).}
\label{slitoddsep}
\end{eqn}

\begin{figure}[hbt!]\centering{\includegraphics[width=.47\textwidth]{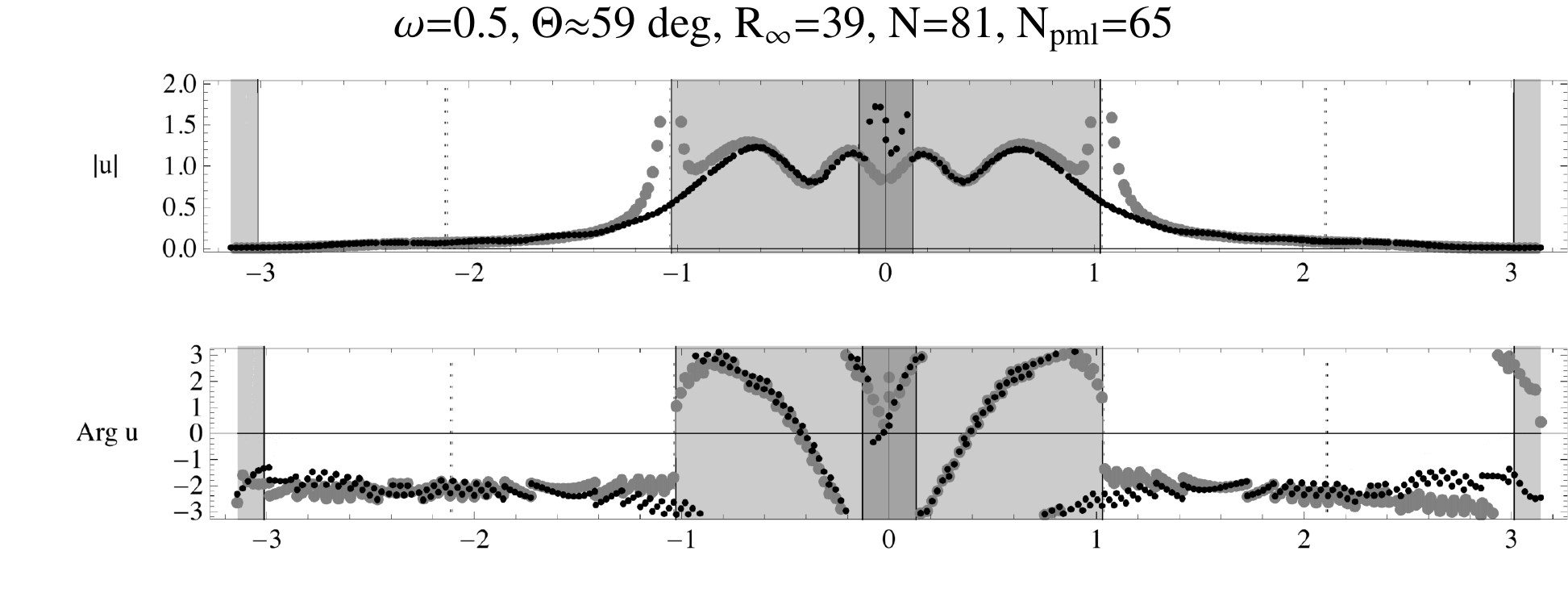}\includegraphics[width=.47\textwidth]{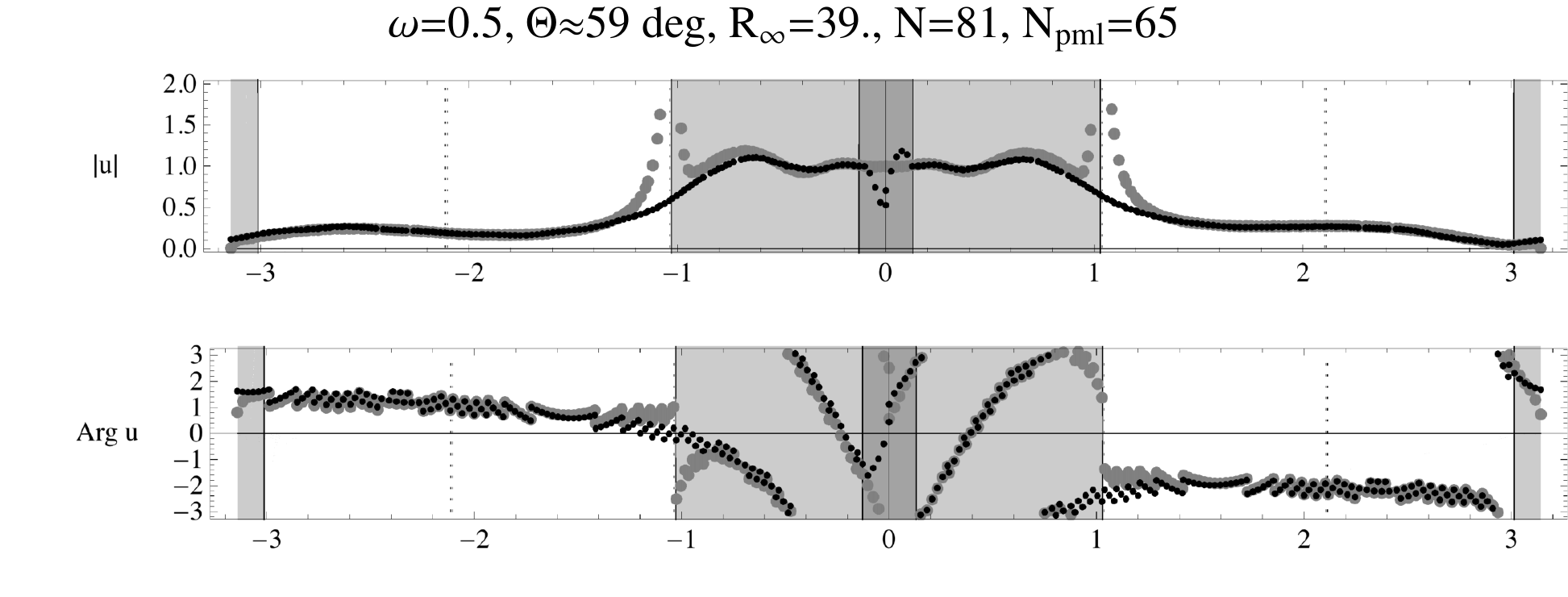}\\}{(a)\includegraphics[width=.47\textwidth]{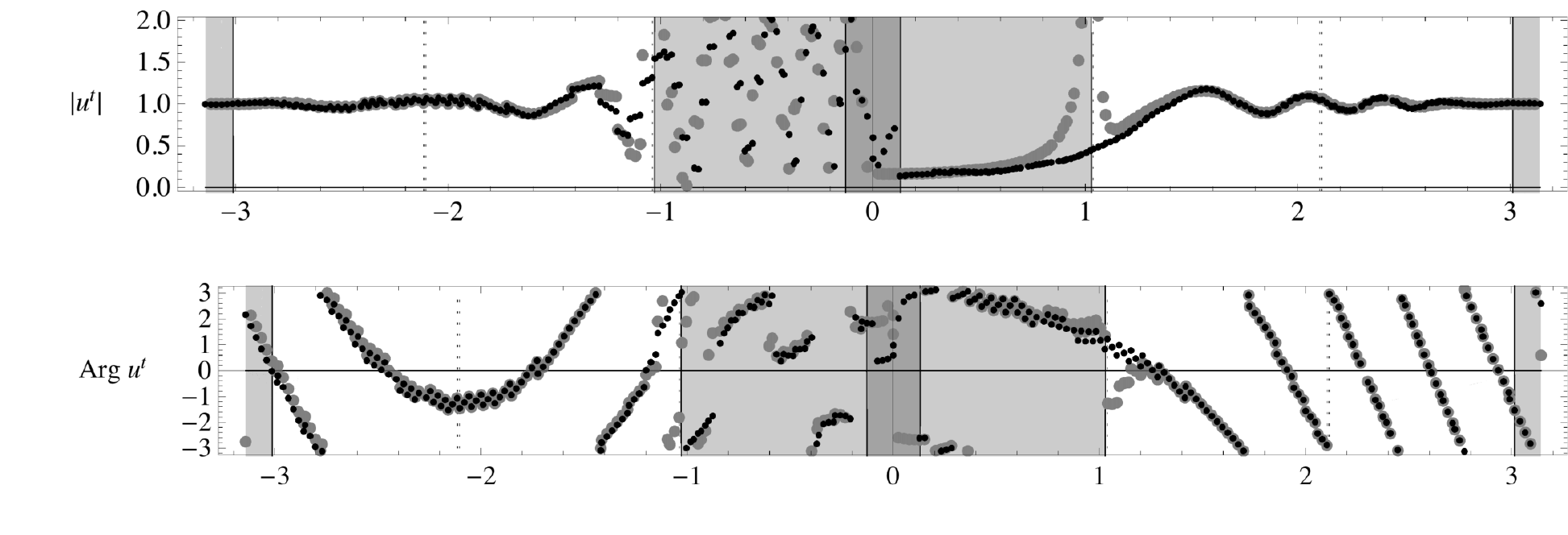}(b)\includegraphics[width=.47\textwidth]{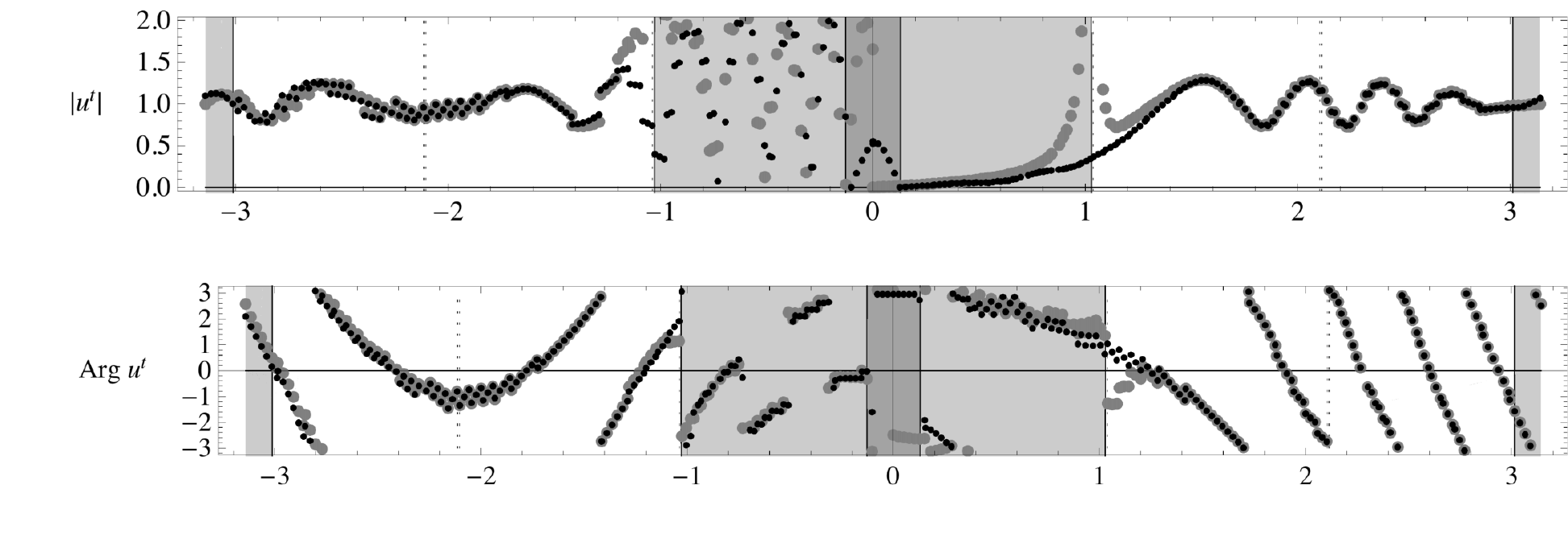}}\caption{Comparison between asymptotic approximation {(gray dots)} and numerical solution {(black dots)} for the scattered and total field for square lattice with a pair of semi-infinite (a) cracks and (b) rigid constraints. Here ${N_w}=2{N}-1=9$ and $N_{\text{grid}}=81, N_{\text{pml}}=65.$}\label{utot_AbsArg_parity_0_w_5_Theta_59_Nt_5_defecttype_12}\end{figure}

\begin{figure}[hbt!]
\centering
{\includegraphics[width=.47\textwidth]{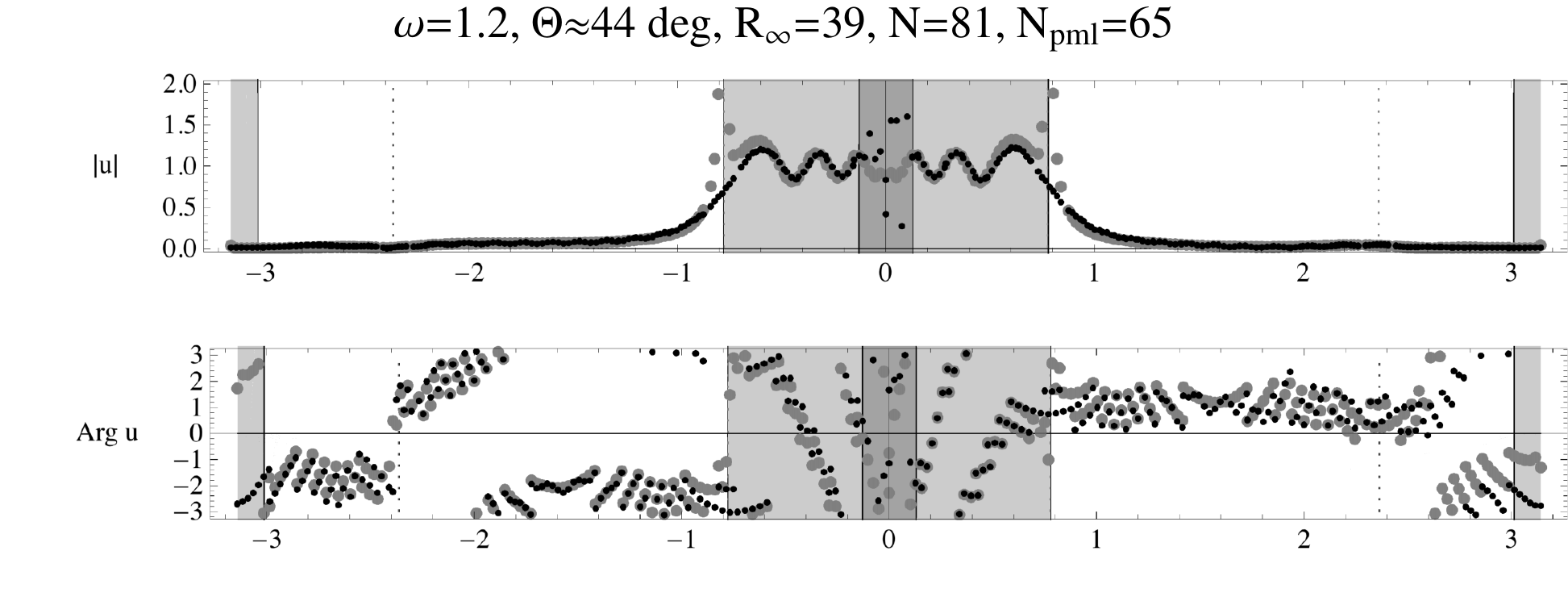}\includegraphics[width=.47\textwidth]{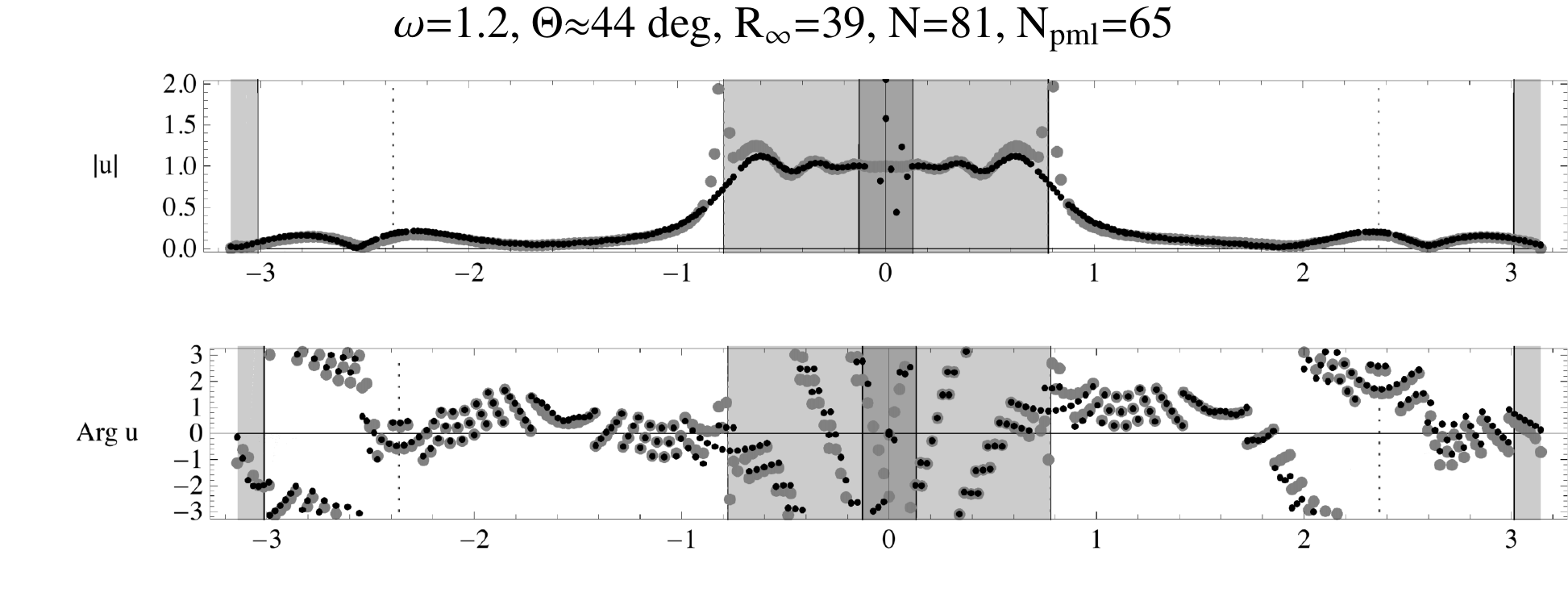}\\
(a)\includegraphics[width=.47\textwidth]{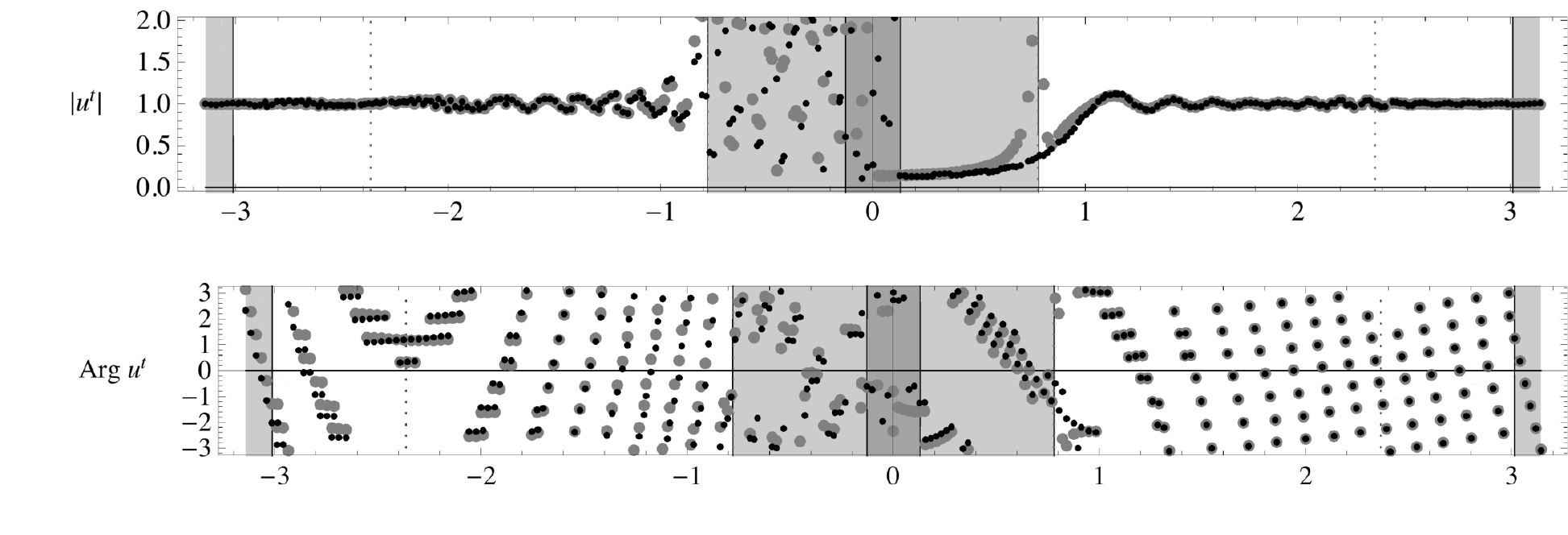}(b)\includegraphics[width=.47\textwidth]{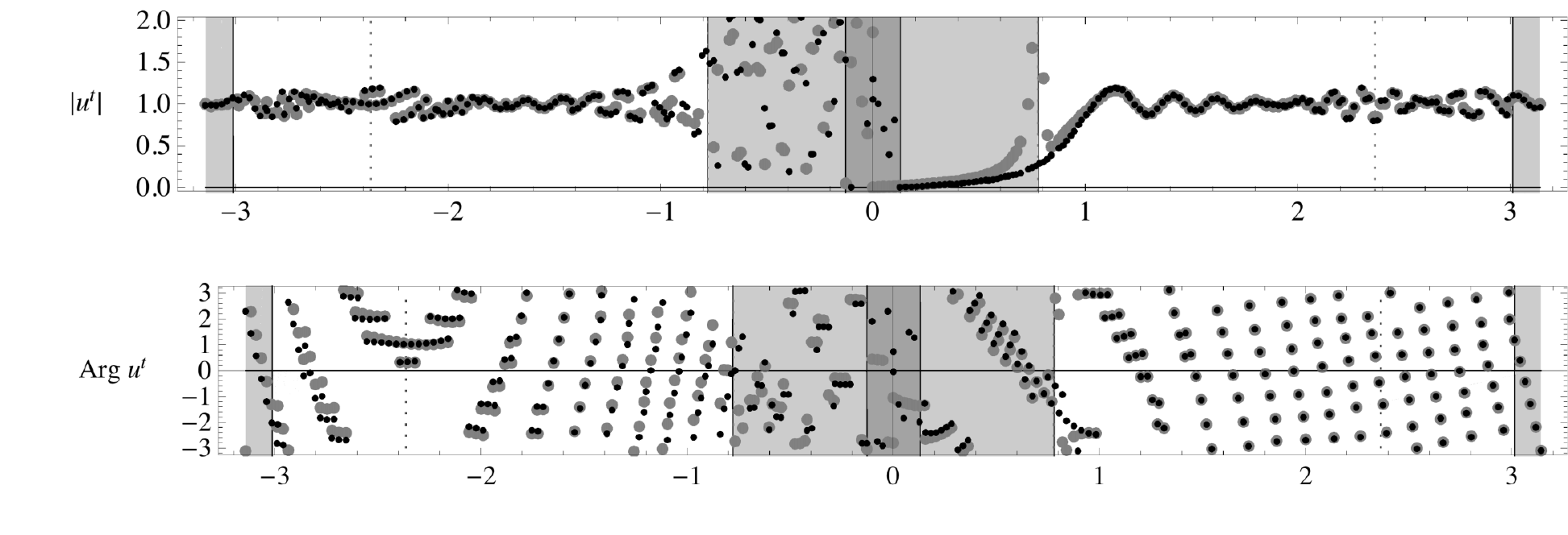}}
\caption{Same as Fig. \ref{utot_AbsArg_parity_0_w_5_Theta_59_Nt_5_defecttype_12} except for incident wave parameters.
}
\label{utot_AbsArg_parity_0_w_12_Theta_44_Nt_5_defecttype_12}
\end{figure}

\begin{figure}[hbt!]
\centering
{\includegraphics[width=.47\textwidth]{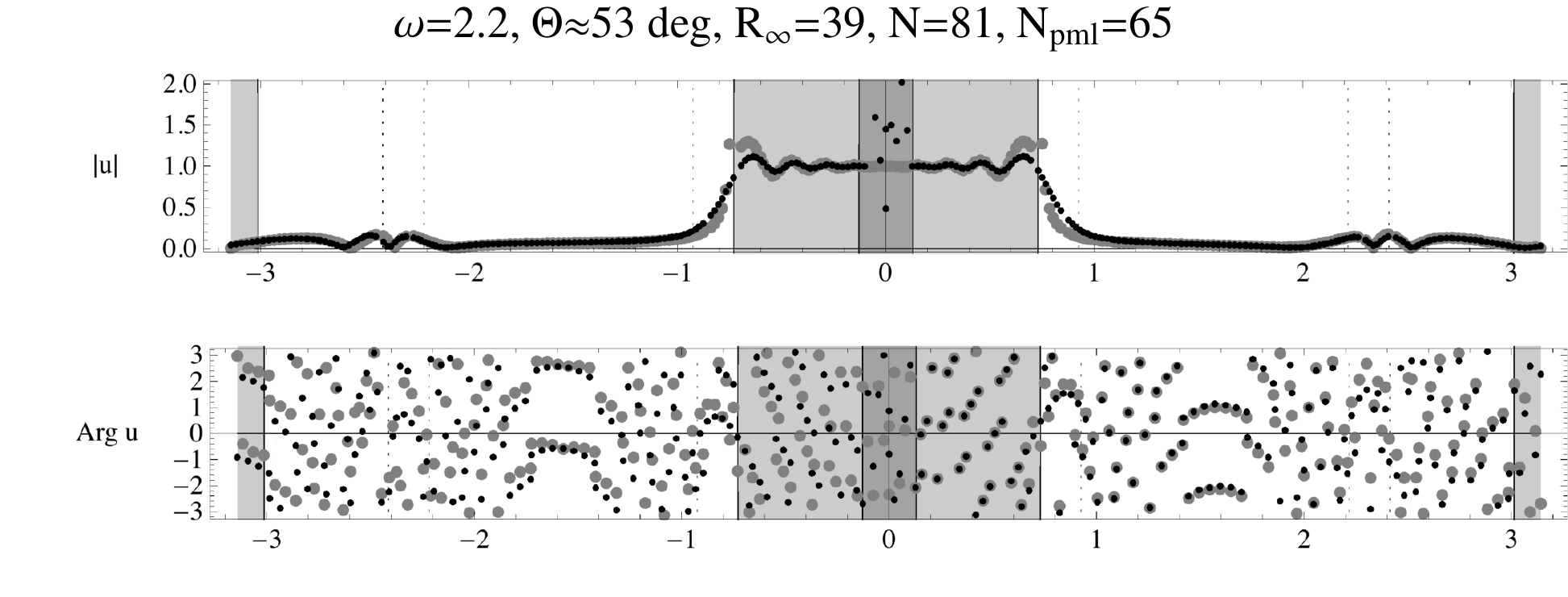}\includegraphics[width=.47\textwidth]{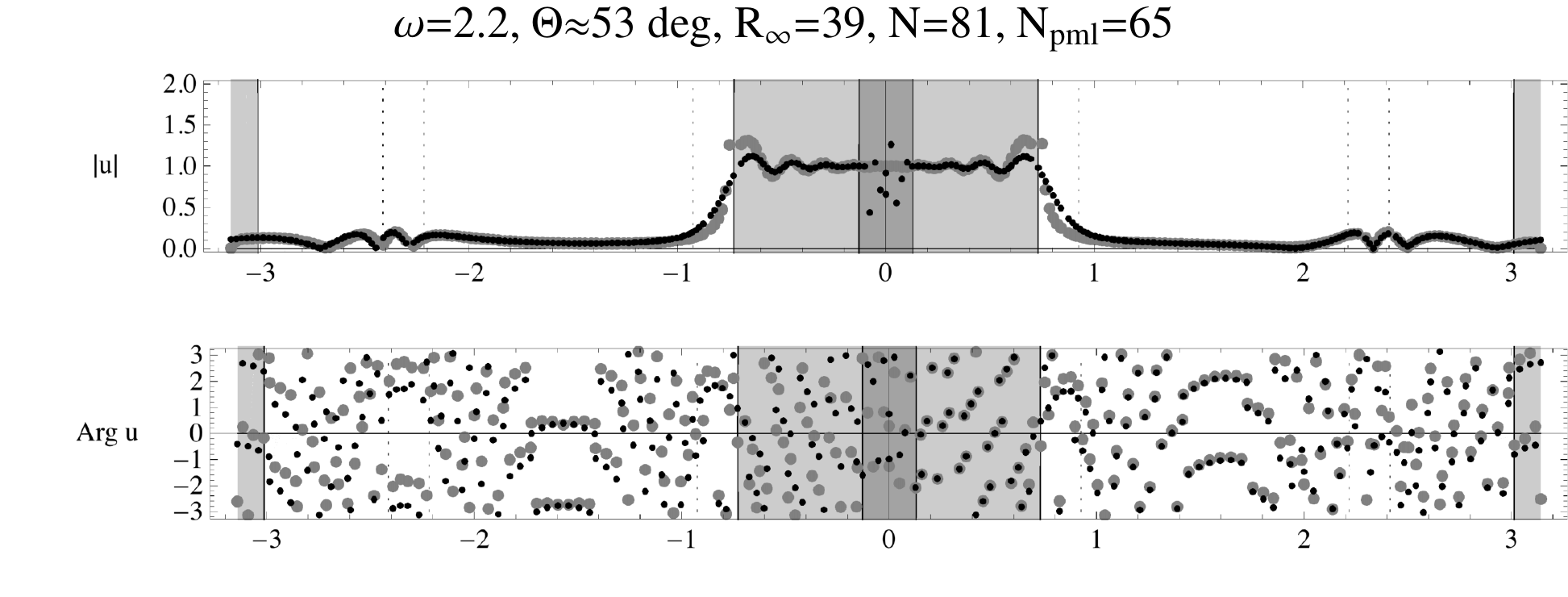}\\
(a)\includegraphics[width=.47\textwidth]{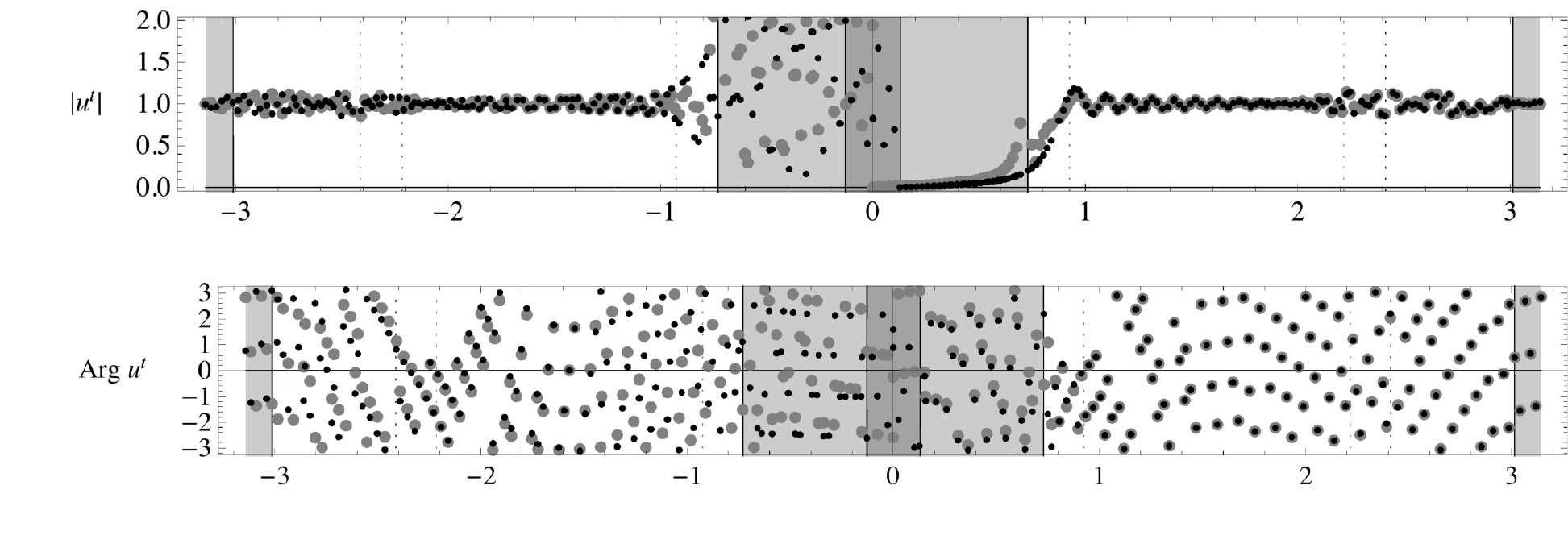}(b)\includegraphics[width=.47\textwidth]{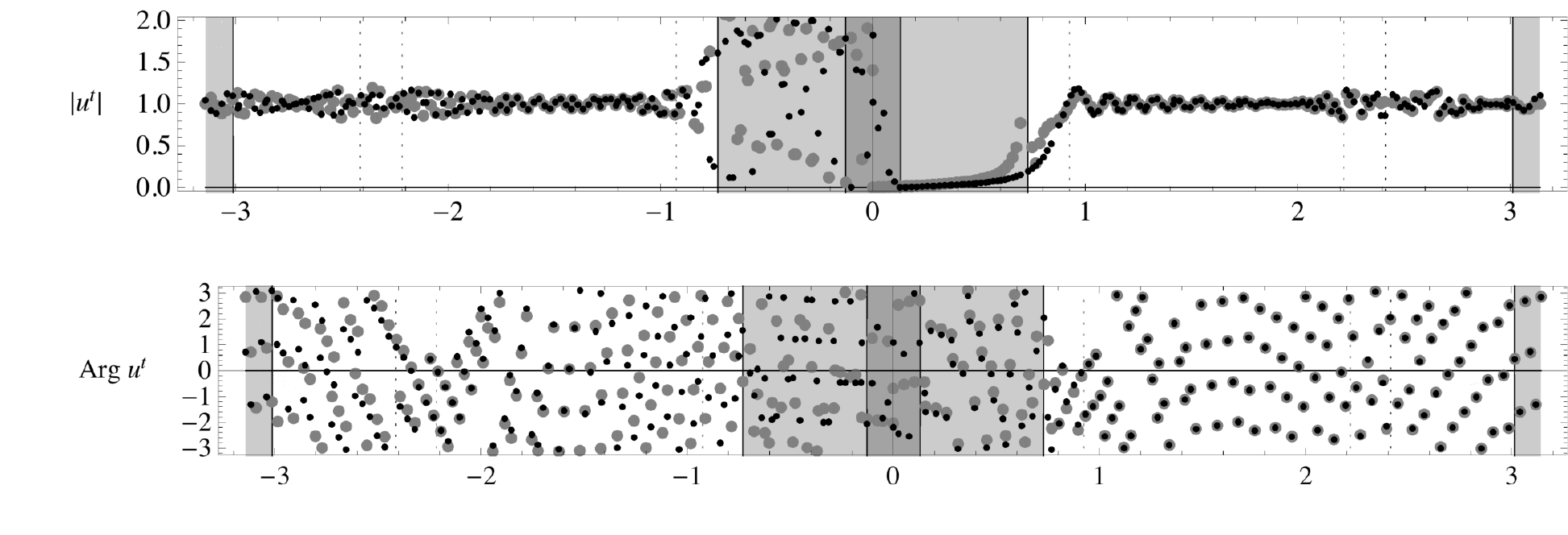}}
\caption{Same as Fig. \ref{utot_AbsArg_parity_0_w_5_Theta_59_Nt_5_defecttype_12} except for incident wave parameters.
}
\label{utot_AbsArg_parity_0_w_22_Theta_53_Nt_5_defecttype_12}
\end{figure}

The construction by superposition provided in this section can be used to obtain the far-field approximation in conjunction with the expressions derived in \S\ref{farfield}. 
The results based on numerical scheme (summarized in Appendix of \cite{Bls0}) and far-field asymptotics have been found to 
coincide in a manner similar to single defect \cite{Bls0,Bls1}. 
Some illustrative results are presented in Fig. \ref{utot_AbsArg_parity_0_w_5_Theta_59_Nt_5_defecttype_12}--Fig. \ref{utot_AbsArg_parity_0_w_22_Theta_53_Nt_5_defecttype_12} {where the modulus and argument of the scattered as well as total displacement field have been plotted relative to the angle ${\theta}$ (on the horizontal axis) and for a fixed (approximate) circle of radius ${R}=39$ according to the polar coordinates \eqref{polar_sq}. The numerical solution is based on a scheme, summarized in an Appendix of \cite{Bls0}, with $N_{\text{grid}}=81, N_{\text{pml}}=65$ (same as that stated in the caption of Fig. \ref{utot_parity_0_w_5_Theta_59_Nt_5_defecttype_12}).}

\section{Concluding remarks}
In this paper, an analysis of a discrete analogue of diffraction by a pair of semi-infinite cracks or rigid constraints is presented following the analysis of \cite{Heins1,Heins2}.
The exact solution is obtained by the discrete {{WH}} method. 
An asymptotic approximation of the exact solution in far field, away from the region corresponding to the proximity of pole and saddle, agrees with the numerical solution as well. 
An illustrative calculation of the near-tip field is carried out as the closed form expressions for the the first broken-bond length, in any of the two cracks, and the displacement of a site adjacent to the rigid constraint tip, are presented.

It is easy to see that there are certain limiting cases of the studied structure leading to interesting configurations; for example, a single semi-infinite defect, as well as a surface step with possibly mixed boundary condition.

As the separation ${N}\to\infty$, naturally, the {field near one of the tip of the two defects in} two crack or two constraint problem reduces to that of a discrete Sommerfeld problem \cite{Bls0,Bls1}.
When ${N}=1,$ for the odd separation the problem {again becomes a discrete Sommerfeld problem \cite{Bls0,Bls1}}, while for the even separation, it {is a case of scattering due to the presence of ``double" crack or ``double" constraint}.

Further, within in the {geometrically} reduced diffraction problem {on the lattice half-plane}, 
a limiting case coincides with that studied recently \cite{Bls10mixed}. 
When ${N}=1$ but ${{\upbeta}}=0, {{\upgamma}}=0$, the problem reduces to that for a variant of mixed boundary condition at ${\mathtt{y}}=1$.
With respect to Fig. \ref{squarelattice_modeIII_singledefects_halfplane}(a), when ${N}=1$ and ${{\upbeta}}=0, {{\upgamma}}=-1$, the problem reduces to that for a single step on a free surface.
With respect to Fig. \ref{squarelattice_modeIII_singledefects_halfplane}(b), when ${N}=1$ and ${{\upbeta}}=0, {{\upgamma}}=0$, the problem reduces to that for a single step on a fixed surface.
On the other hand, when ${N}=1$ but ${{\upbeta}}=0, {{\upgamma}}=-1$, the problem reduces to that for a variant of mixed boundary condition at ${\mathtt{y}}=0$.

In place of the infinite square lattice, if the pair of semi-infinite defects are placed symmetrically on a square lattice waveguide, then the same formulation can be extended to what are known as {\em trifurcated} waveguides \cite{PaceMittra,MahRawlins}. The additional confinement induces different structure factors in the two {{WH}} kernels. The exact solution can be easily arrived at and closed form expressions for the transmission problem can be found;
it is useful to recall the analysis of \cite{Bls9s} as the reflectance and transmittance of the junction can be constructed. 
More pertinent from the viewpoint of transport is the scattering matrix which {has been found to} admit a succinct expression as well, the presentation of which {in the public domain} has been deferred.

{Last but not the least, there remains an issue of the continuum limit. For the considered case of positive imaginary part of ${\upomega}$, it is left as an exercise (one possibility involves the tools that are used in \cite{Bls31}) to prove that the low frequency limit (i.e. with $\la\to0$ but fixed $\icf$ and ${N}\la$; recall ${\upomega}=\la\icf$) coincides with that of the well known solution \cite{Heins1,Heins2} provided the separation between the semi-infinite defects ${N}$ scales naturally as $1/\la$. An interesting non-trivial question is the rigorous statement and proof of the counterpart corresponding to ${\upomega}_2=0$? Note that the same question remains open for the discrete Sommerfeld problems as well \cite{Bls0, Bls1,Bls2,Bls3,Bls31}. In the same vein, another curious question concerns the scattering problem involving parallel defects (on the square lattice) and its solution as ${\upomega}_1\to2$ or ${\upomega}_1\to2\sqrt{2}$?}

\section*{Acknowledgement}
BLS acknowledges the partial support of SERB MATRICS grant $MTR/2017/000013.$
GM acknowledges MHRD (India) and IITK for providing financial assistance in the form of Senior Research Fellowship.

\printbibliography

\begin{appendix}

\section{Discrete Fourier transform}
\label{wellposedness}
Akin to \cite{Bls0}, {in case of bulk incidence \eqref{uinc}}, it can be easily shown that {${\su}_{{\mathtt{y}}}^F$, given by \eqref{discreteFT},}
is analytic inside the annulus
${{\mathscr{A}}}_u{:=}\{{{z}}\in{\mathbb{C}}: {{\mathrm{R}}}_+< |{{z}}|< {{\mathrm{R}}}_-\},$ 
where
${{\mathrm{R}}}_+=e^{-{\upkappa}_2}, {{\mathrm{R}}}_-=e^{{\upkappa}_2\cos{\Theta}}$ (for $0\le{\mathtt{y}}\le{N}$, in fact, ${{\mathrm{R}}}_-=e^{+{\upkappa}_2}$)
Based on above discussion, the discrete Fourier transform ${\su}_{{\mathtt{y}}}^F$ of the sequence $\{{\su}_{{\mathtt{x}}, {\mathtt{y}}}\}_{{\mathtt{x}}\in{\mathbb{Z}}}$ is well defined for all ${\mathtt{y}}\in{\mathbb{Z}}.$ 
Using the discrete Fourier transform 
{\eqref{discreteFT}}, 
the general solution {of the scattered wave field according to} the discrete Helmholtz equation, {i.e., \eqref{dHelmholtz} with ${\su}^{\totwave}$ replaced by ${\su}$ since ${\su}^{\inc}$ automatically satisfies it}, 
is given by the expression
${\su}_{{\mathtt{y}}}^F={\mathrm{c}}_1{{\lambda}}^{{\mathtt{y}}}+{\mathrm{c}}_2{{\lambda}}^{-{\mathtt{y}}}, $
where ${\mathrm{c}}_1, {\mathrm{c}}_2$ are arbitrary analytic functions on ${{\mathscr{A}}}$
and
the function ${{\lambda}}$ is defined by \cite{Bls0,Bls1,Slepyanbook}, 
\beqan
{{\lambda}}{:=}
\frac{\sr-\sh}{\sr+\sh},
\text{ on }{\mathbb{C}}\setminus{\mathscr{B}}, \quad
\label{lambdadef}
\text{where }
{\sh}{:=}\sqrt{{\sH}}, {\sr}{:=}\sqrt{{\sR}}, \\
\text{with }
{\sH}{:=}{\sQ}-2, {\sR}{:=}{\sQ}+2, 
{{z}}\in{\mathbb{C}},
\label{defh2}
\eeqan
and ${\mathscr{B}}$ as the union of branch cuts for ${{\lambda}}$ 
borne out of the chosen branch 
for ${\sh}$ and ${\sr}$ such that 
$|{{\lambda}}({{z}})|\le1, {{z}}\in{\mathbb{C}}\setminus{\mathscr{B}}.$
Following \cite{Bls0}, lan annulus ${{\mathscr{A}}}\subset{\mathbb{C}}$ is defined by
\begin{eqn}
{{\mathscr{A}}}{:=}{{\mathscr{A}}}_u\cap{{\mathscr{A}}}_L, ~{{\mathscr{A}}}_L{:=}\{{{z}}\in{\mathbb{C}} : {{\mathrm{R}}}_L< |{{z}}|< {{\mathrm{R}}}_L^{-1}\}, {{\mathrm{R}}}_L{:=}\max\{|{{z}}_{\sh}|, |{{z}}_{\sr}|\},
\label{annAAL}
\end{eqn}
where ${{z}}_{\sh}$ and ${{z}}_{\sr}$ are zeros of $\sh$ and $\sr$, respectively.

\clearpage
{\bf Supplementary 1: Expression for ${\su}_{-1,{{N}}}$}\\
{\small Article Title:``Discrete scattering by a pair of parallel defects"\\
Authors: Sharma BL and Maurya G, \\
Journal: Philosophical Transactions of the Royal Society A: Mathematical, Physical and Engineering Sciences}\\

\setcounter{page}{1}

Equation \eqref{uNFeq} implies
${\su}_{{{N}}; -}={\sQ}^{-1}(-{{\mathtt{W}}}_{{N}}+{\su}_{{{N}}+1; -}+{\su}_{{{N}}-1; -})
={\sQ}^{-1}(-{{\mathtt{W}}}_{{N}}+{\mathscrpring}_{{N}; -}).$
Using the inverse discrete Fourier transform {\eqref{discreteFT}}, 
\begin{eqn}
{\su}_{-1,{N}}&=\frac{1}{2\pi i}\oint_{{\mathcal{C}}} {\su}_{{{N}}; -}({{z}}){{z}}^{-1-1}d{{z}}
=\frac{1}{2\pi i}\oint_{{\mathcal{C}}} \frac{-{{\mathtt{W}}}_{{N}}({{z}})+{\mathscrpring}_{{N}; -}({{z}})}{{\sQ}({{z}})}{{z}}^{-2}d{{z}}.
\end{eqn}
With
${\sQ}({{z}})={{z}}_{\sq}^{-1}(1-{{z}}_{\sq}{{z}})(1-{{z}}_{\sq}{{z}}^{-1})=-{{z}}^{-1}({z}-{{z}}_{\sq})({{z}}-{z}_{\sq}^{-1}),$ above implies
\begin{eqn}
{\su}_{-1,{N}}
&=\frac{1}{2\pi i}\oint_{{\mathcal{C}}} \frac{-{{\mathtt{W}}}_{{N}}({{z}})+{\mathscrpring}_{{N}; -}({{z}})}{-({z}-{{z}}_{\sq})({{z}}-{z}_{\sq}^{-1})}{{z}}^{-1}d{{z}}
=({{\mathtt{W}}}_{{N}}({{z}}_{\sq})-{\mathscrpring}_{{N}; -}({{z}}_{\sq})){z}_{\sq}^{-1}+({{\mathtt{W}}}_{{N}}(0)-{\mathscrpring}_{{N}; -}(0))\\
&=({\su}_{-1, {N}}+{{z}}_{\sq} {\su}^{\inc}_{0, {N}}-{\mathscrpring}_{{N}; -}({{z}}_{\sq})){z}_{\sq}^{-1}+({\su}_{-1, {N}}-0),
\end{eqn}
\begin{eqn}
\text{where }{\mathscrpring}_{{N}; -}({{z}}_{\sq})&={{\mathtt{C}}}_-({{z}}_{\sq}){{\mathtt{L}}}_{-}({{z}}_{\sq})\\
&={{\mathtt{L}}}_{-}({{z}}_{\sq})({\su}_{-1, {N}}({{\mathtt{L}}}_{-}^{-1}({{z}}_{\sq})-{\overline{l}}_{-0})+{{z}}_{\sq} {\su}^{\inc}_{0, {N}}({{\mathtt{L}}}_{-}^{-1}({{z}}_{\sq})-{l_{}}_{+0})+\su^{\inc}_{0,{N}}\delta_{D+}({{z}}_{\sq} {{z}}_{{P}}^{-1})\\&\big({\sQ}({{z}}_{\sq}){{\mathtt{L}}}_{-}^{-1}({{z}}_{\sq})-{\sQ}({{z}}_{{P}}){{\mathtt{L}}}_-^{-1}({{z}}_{{P}})+{\overline{l}}_{-0}({{z}}_{\sq}^{-1}-{{z}}_{{P}}^{-1})+{{l}}_{+0}({{z}}_{\sq}-{{z}}_{{P}})\big))\\
&={\su}_{-1, {N}}(1-{{\mathtt{L}}}_{-}({{z}}_{\sq}){\overline{l}}_{-0})+{{z}}_{\sq} {\su}^{\inc}_{0, {N}}(1-{{\mathtt{L}}}_{-}({{z}}_{\sq}){l_{}}_{+0})+\su^{\inc}_{0,{N}}\delta_{D+}({{z}}_{\sq} {{z}}_{{P}}^{-1})\\&\big({\sQ}({{z}}_{\sq})-{{\mathtt{L}}}_{-}({{z}}_{\sq}){\sQ}({{z}}_{{P}}){{\mathtt{L}}}_-^{-1}({{z}}_{{P}})+{\overline{l}}_{-0}({{z}}_{\sq}^{-1}-{{z}}_{{P}}^{-1}){{\mathtt{L}}}_{-}({{z}}_{\sq})+{{l}}_{+0}({{z}}_{\sq}-{{z}}_{{P}}){{\mathtt{L}}}_{-}({{z}}_{\sq})\big).\notag
\end{eqn}
\begin{eqn}
\text{Hence, }
0&={\su}_{-1, {N}}(-{{\mathtt{L}}}_{-}({{z}}_{\sq}){\overline{l}}_{-0})+{{z}}_{\sq} {\su}^{\inc}_{0, {N}}(-{{\mathtt{L}}}_{-}({{z}}_{\sq}){l_{}}_{+0})+\su^{\inc}_{0,{N}}\frac{{{z}}_{\sq}}{{{z}}_{\sq}-{{z}}_{{P}}}\\&\big({\sQ}({{z}}_{\sq})-{{\mathtt{L}}}_{-}({{z}}_{\sq}){\sQ}({{z}}_{{P}}){{\mathtt{L}}}_-^{-1}({{z}}_{{P}})+{\overline{l}}_{-0}({{z}}_{\sq}^{-1}-{{z}}_{{P}}^{-1}){{\mathtt{L}}}_{-}({{z}}_{\sq})+{{l}}_{+0}({{z}}_{\sq}-{{z}}_{{P}}){{\mathtt{L}}}_{-}({{z}}_{\sq})\big)\\
&=-{\su}_{-1, {N}}{{\mathtt{L}}}_{-}({{z}}_{\sq}){\overline{l}}_{-0}-\frac{\su^{\inc}_{0,{N}}{{z}}_{\sq}}{{{z}}_{\sq}-{{z}}_{{P}}}{{\mathtt{L}}}_{-}({{z}}_{\sq}){\sQ}({{z}}_{{P}}){{\mathtt{L}}}_-^{-1}({{z}}_{{P}})-{\overline{l}}_{-0}{{z}}_{{P}}^{-1}{{\mathtt{L}}}_{-}({{z}}_{\sq})\su^{\inc}_{0,{N}},\notag
\end{eqn}
which gives
${\su}_{-1, {N}}=\su^{\inc}_{0,{N}}(-\frac{{{z}}_{\sq}}{{{z}}_{\sq}-{{z}}_{{P}}}\frac{{\sQ}({{z}}_{{P}})}{{\overline{l}}_{-0}{{\mathtt{L}}}_-({{z}}_{{P}})}-{{z}}_{{P}}^{-1}),$
i.e., \eqref{un1Ntot} holds.
Similarly, in the case of {\em incidence from the waveguide}, \eqref{CpmC_altinc} implies
\begin{eqn}
{{\mathtt{C}}}_{-}({{z}})&={\su}_{-1, {N}}({{\mathtt{L}}}_-^{-1}({{z}})-{\overline{l}}_{-0})-{\su}^{\inc}_{0,{N}-1}\delta_{D-}({{z}} {{z}}_{{P}}^{-1})
\big({{\mathtt{L}}}_{-}^{-1}({{z}})-{{\mathtt{L}}}_+({{z}}_{{P}})\big),
\end{eqn}
so that by \eqref{uNFeq_altinc},
\begin{eqn}
{\su}_{-1,{N}}&=\frac{1}{2\pi i}\oint_{{\mathcal{C}}} \frac{-{{\mathtt{W}}}_{{N}}({{z}})+{\mathscrpring}_{{N}; -}({{z}})+{\mathscrpring}^{\inc}_{{N}; -}({{z}})}{-{{z}}^{-1}({z}-{{z}}_{\sq})({{z}}-{z}_{\sq}^{-1})}{{z}}^{-2}d{{z}}\\
&=({{\mathtt{W}}}_{{N}}({{z}}_{\sq})-{\mathscrpring}_{{N}; -}({{z}}_{\sq})-{\mathscrpring}^{\inc}_{{N}; -}({{z}}_{\sq})){z}_{\sq}^{-1}/({{z}}_{\sq}-{z}_{\sq}^{-1})+({{\mathtt{W}}}_{{N}}(0)-{\mathscrpring}_{{N}; -}(0)-{\mathscrpring}^{\inc}_{{N}; -}(0))\\
&=({\su}_{-1, {N}}+{{z}}_{\sq} {\su}^{\inc}_{0, {N}}-{\mathscrpring}_{{N}; -}({{z}}_{\sq})-{\mathscrpring}^{\inc}_{{N}; -}({{z}}_{\sq})){z}_{\sq}^{-1}/({{z}}_{\sq}-{z}_{\sq}^{-1})+({\su}_{-1, {N}}-0),
\end{eqn}
\begin{eqn}
\text{where }
{\mathscrpring}_{{N}; -}({{z}}_{\sq})&={{\mathtt{C}}}_-({{z}}_{\sq}){{\mathtt{L}}}_{-}({{z}}_{\sq})\\
&={{\mathtt{L}}}_{-}({{z}}_{\sq})({\su}_{-1, {N}}({{\mathtt{L}}}_-^{-1}({{z}}_{\sq})-{\overline{l}}_{-0})-{\su}^{\inc}_{0,{N}-1}\delta_{D-}({{z}}_{\sq} {{z}}_{{P}}^{-1})
\big({{\mathtt{L}}}_{-}^{-1}({{z}}_{\sq})-{{\mathtt{L}}}_+({{z}}_{{P}})\big))\\
&={\su}_{-1, {N}}(1-{{\mathtt{L}}}_{-}({{z}}_{\sq}){\overline{l}}_{-0})-\su^{\inc}_{0,{N}-1}\delta_{D-}({{z}}_{\sq} {{z}}_{{P}}^{-1})\big(1-{{\mathtt{L}}}_{-}({{z}}_{\sq}){{\mathtt{L}}}_+({{z}}_{{P}})\big)).
\notag
\end{eqn}
Hence,
$0=-{\su}_{-1, {N}}(-{{\mathtt{L}}}_{-}({{z}}_{\sq}){\overline{l}}_{-0})-\su^{\inc}_{0,{N}-1}\frac{{{z}}_{\sq}}{{{z}}_{\sq}-{{z}}_{{P}}}+\su^{\inc}_{0,{N}-1}\frac{{{z}}_{\sq}}{{{z}}_{\sq}-{{z}}_{{P}}}(1-{{\mathtt{L}}}_{-}({{z}}_{\sq}){{\mathtt{L}}}_+({{z}}_{{P}})),$
which gives
${\su}_{-1, {N}}=-\su^{\inc}_{0,{N}-1}\frac{{{z}}_{\sq}}{{{z}}_{\sq}-{{z}}_{{P}}}\frac{{{\mathtt{L}}}_+({{z}}_{{P}})}{{\overline{l}}_{-0}},$
i.e., \eqref{un1Ntot_altinc} holds.

\end{appendix}

\end{document}